\documentclass[12pt,epsf,amssymb]{article} \textheight =23 truecm
\def\lsim{\raise0.3ex\hbox{$<$\kern-0.75em\raise-1.1ex\hbox{$\sim$}}}
\def\gsim{\raise0.3ex\hbox{$>$\kern-0.75em\raise-1.1ex\hbox{$\sim$}}}
\def\noi{\noindent} \def\nn{\nonumber} \def\bea{\begin{eqnarray}}
\def\eea{\end{eqnarray}} \def\beq{\begin{equation}}
\def\eeq{\end{equation}} 
\def\beeq{\begin{eqnarray}} \def\eeeq{\end{eqnarray}} \def\R{ {\rm R
\kern -.31cm I \kern .15cm}} \def\C{ {\rm C \kern -.15cm \vrule
width.5pt \kern .12cm}} \def\Z{ {\rm Z \kern -.27cm \angle \kern
.02cm}} \def\N{ {\rm N \kern -.26cm \vrule width.4pt \kern .10cm}}
\def\1{{\rm 1\mskip-4.5mu l} }

\usepackage{graphics} \usepackage{graphicx} \usepackage{epsfig}

\begin{document} \begin{center} {\large \bf Lagrangian perturbations at order 1/m$_{\bf Q}$ and the non-forward amplitude in Heavy Quark Effective Theory} \\

\vskip 1 truecm {\bf F. Jugeau, A. Le Yaouanc, L. Oliver and J.-C.
Raynal}\\

{\it Laboratoire de Physique Th\'eorique}\footnote{Unit\'e Mixte de
Recherche UMR 8627 - CNRS }\\    {\it Universit\'e de Paris XI,
B\^atiment 210, 91405 Orsay Cedex, France} \end{center}

\vskip 1 truecm

\begin{abstract}  
We pursue the program of the study of the non-forward amplitude in
HQET. We obtain new sum rules involving the elastic subleading form
factors $\chi_i(w)$ $(i = 1,2, 3)$ at order $1/m_Q$ that originate from
the ${\cal L}_{kin}$ and ${\cal L}_{mag}$ perturbations of the
Lagrangian. To obtain these sum rules we use two methods. On the one
hand we start simply from the definition of these subleading form
factors and, on the other hand, we use the Operator Product Expansion.
To the sum rules contribute only the same intermediate states $\left (
j^P, J^P\right ) = \left ( {1 \over 2}^-, 1^-\right ) , \left ( {3\over
2}^-, 1^-\right )$ that enter in the $1/m_Q^2$ corrections of the axial
form factor $h_{A_1}(w)$ at zero recoil. This allows to obtain a lower
bound on $- \delta_{1/m^2}^{(A_1)}$ in terms of the $\chi_i(w)$ and the
shape of the elastic IW function $\xi (w)$. We find also lower bounds
on the $1/m_Q^2$ correction to the form factors $h_+(w)$ and $h_1(w)$ at zero recoil. An important theoretical implication is that $\chi '_1(1)$, $\chi_2(1)$ and $\chi ' _3(1)$ ($\chi_1(1) = \chi_3(1) = 0$ from Luke theorem)  must vanish when the slope and the curvature attain their
lowest values $\rho^2 \to {3 \over 4}$, $\sigma^2 \to {15 \over 16}$. We discuss possible implications
on the precise determination of $|V_{cb}|$. \end{abstract}

\vskip 2 truecm

\noi LPT Orsay 05-43 \par \noi July 2005
\par \vskip 1 truecm

\noindent e-mails : frederic.jugeau@th.u-psud.fr,
leyaouan@th.u-psud.fr, oliver@th.u-psud.fr \newpage \pagestyle{plain}

\section{Introduction.} \hspace*{\parindent} 
The study of the non-forward amplitude, proposed first by Uraltsev
\cite{1r}
\beq
\label{1e}
T_{fi}(q) = i \int d^4x\ e^{-iq\cdot x} \ < B(v_f)|T[J_f(0) J_i(x)]|B(v_i)>
\eeq

\noi where $v_i$ is in general different from $v_f$ and
\beq
\label{2e}
J_f(0) = \overline{b} (0) \Gamma_f c(0) \qquad J_i(x) = \overline{c} (x) \Gamma_i b(x)
\eeq

\noi ($\Gamma_i$, $\Gamma_f$ are arbitrary Dirac matrices) has been
very fruitful in Heavy Quark Effective Theory (HQET). \par

In the heavy quark limit, sum rules (SR) that generalize Bjorken
\cite{2r} and Uraltsev \cite{1r} SR have been obtained within the Operator Product Expansion (OPE) that yield to
bounds for all derivatives of the elastic Isgur-Wise (IW) function $\xi (w)$
\cite{3r} \cite{4r}, in particular for the curvature \cite{5r}. The
radiative corrections to these SR and bounds in the framework of HQET
have been computed by Dorsten \cite{6r}. \par

In a recent paper we have extended our formalism to the subleading
order in $1/m_Q$ \cite{7r}. We did obtain the interesting relations,
{\it valid for all $w$}~: 
\beq
\label{3e}
\overline{\Lambda}\xi (w) = 2(w+1) \sum_n \Delta E_{3/2}^{(n)} \tau_{3/2}^{(n)} (1) \tau_{3/2}^{(n)}(w) + 2 \sum_n \Delta E_{1/2}^{(n)} \tau_{1/2}^{(n)}(1) \tau_{1/2}^{(n)}(w)
 \eeq 
\beq
\label{4e}
\xi_3 (w) = (w+1) \sum_n \Delta E_{3/2}^{(n)} \tau_{3/2}^{(n)} (1) \tau_{3/2}^{(n)}(w) - 2 \sum_n \Delta E_{1/2}^{(n)} \tau_{1/2}^{(n)}(1) \tau_{1/2}^{(n)}(w) \ .
 \eeq 

These remarkably simple relations were the basic results of ref.
\cite{7r}. Both subleading quantities $\overline{\Lambda} \xi (w)$ and
$\xi_3(w)$ can be expressed in terms of the leading quantities, namely
IW functions $\tau_j^{(n)}(w)$ and level spacings $\Delta E_j^{(n)}$
$\left (j = {1 \over 2}, {3 \over 2} \right )$. These equations give
information on the $1/m_Q$ {\it Current} perturbations to the matrix
elements. In the present
paper we will deal with the {\it Lagrangian} perturbations. \par

The paper is organized as follows. Section 2 gives a simple derivation
of the relevant SR, starting from the definition of the different
subleading Lagrangian form factors. In Section 3 we summarize the basic
results and comment on general theoretical features of the SR. In
Section 4 we recall the contribution of $1^-$ intermediate states to
the OPE sum rule at zero recoil at order $1/m_Q^2$ for the form factor
$B \to D^*$. In Section 5, using Schwarz inequality, we obtain a bound
on the correction $\delta_{1/m^2}$ to $F_{B\to D^*}(1)$ in terms of the
Lagrangian elastic subleading form factors and the elastic Isgur-Wise
function. In Section 6 we also obtain lower bounds on the $1/m_Q^2$ corrections to the form factors $h_+(w)$ and $h_1(w)$ at $w = 1$. In
Section 7 we summarize some theoretical features of the obtained
bounds. In Section 8  we demonstrate that $\chi '_1(1)$, $\chi_2(1)$ and $\chi '_3(1)$ must vanish in the limit in which the slope $\rho^2$ and curvature $\sigma^2$ of the elastic IW function $\xi (w)$ attain their lowest values. In Section 9 we discuss
phenomenological implications of our results for the exclusive
determination of $|V_{cb}|$ and in Section 10 we conclude. In Appendix
A we derive the same SR as in Section 2 using the Operator Product
Expansion (OPE), following the same method developed for the derivation
of the Current SR in \cite{7r}. In Appendix B we make a numerical analysis of the obtained bounds and in Appendix C we discuss the radiative corrections. 

\section{New sum rules on Lagrangian perturbations.} \hspace*{\parindent}
In this section, we will formulate new SR for the {\it Lagrangian}
perturbations, parallel to the ones on the {\it Current} perturbations
(\ref{3e})-(\ref{4e}). \par

Instead of using the OPE, we will here simply use the definition of the
subleading elastic ${1\over 2}^- \to {1 \over 2}^-$ functions
$\chi_i(w)$ $(i = 1,2,3)$ \cite{18r}
\bea
\label{5e}
&&<D(v') |i \int dxT [J^{cb} (0), {\cal L}_v^{(b)}(x)]|B(v)>\ = \nn \\
&&{1 \over 2m_b} \left \{ - 2 \chi_1(w) Tr \left [\overline{D}(v') \Gamma B(v)\right ] + {1 \over 2} Tr\left  [ A_{\alpha\beta}(v,v') \overline{D}(v') \Gamma P_+ i \sigma^{\alpha\beta} B(v)\right ] \right \}
\eea
\bea
\label{6e}
&&<D(v') |i \int dxT [J^{cb} (0), {\cal L}_{v'}^{(c)}(x)]|B(v)>\ = \nn \\
&&{1 \over 2m_c} \left \{ - 2 \chi_1(w) Tr \left [\overline{D}(v') \overline{\Gamma} B(v)\right ] - {1 \over 2} Tr\left  [ \overline{A}_{\alpha\beta}(v',v) \overline{D}(v') i \sigma^{\alpha\beta} P'_+ \overline{\Gamma} B(v)\right ] \right \}
\eea

\noi with
\bea
\label{7e}
&&A_{\alpha \beta}(v,v') = - 2 \chi_2 (w) \left ( v'_{\alpha} \gamma_{\beta} - v'_{\beta} \gamma_{\alpha}\right ) + 4 \chi_3 (w) i \sigma_{\alpha \beta} \nn \\
&&\overline{A}_{\alpha \beta}(v',v) = - 2 \chi_2 (w) \left ( v_{\alpha} \gamma_{\beta} - v_{\beta} \gamma_{\alpha}\right ) - 4 \chi_3 (w) i \sigma_{\alpha \beta} 
\eea

\noi where $\overline{A} = \gamma^0 A^+\gamma^0$ denotes the Dirac
conjugate matrix, the current $J^{cb}(0)$ denotes
\beq
\label{8e}
J^{cb} = \overline{h}_{v'}^{(c)} \Gamma h_v^{(b)}
\eeq

\noi where $\Gamma$ is any Dirac matrix, and ${\cal L}_v^{(Q)}(x)$ is given by 
\beq
\label{9e}
{\cal L}_v^{(Q)} = {1 \over 2 m_Q} \left [ O_{kin, v}^{(Q)} + O_{mag, v}^{(Q)}\right ]
\eeq

\noi with
\beq
\label{10e}
O_{kin, v}^{(Q)} = \overline{h}_v^{(Q)}(iD)^2 h_v^{(Q)} \qquad O_{mag, v}^{(Q)} = {g_s \over 2} \overline{h}_v^{(Q)} \sigma_{\alpha \beta} G^{\alpha\beta} h_v^{(Q)} \ .
\eeq

In relations (\ref{5e})-(\ref{7e}), the $\chi_i(w)$ $(i= 1, 2, 3)$ have dimensions of mass, and correspond to the definition given by Luke \cite{10r}.

We will now insert intermediate states in the $T$-products (\ref{5e}).
We can separately consider ${\cal L}_{kin}^{(b)}$ or ${\cal
L}_{mag}^{(b)}$. The possible $Z$-diagrams involving heavy quarks
contributing to the $T$-products are suppressed by the heavy quark mass
since they are $b\overline{c}c$ intermediate states.\par

Conveniently choosing the initial and final states, we find the
following results (we use the normalization of the states as made explicit for example in formula (5.6) of ref. \cite{8r})~:\par

(1) With ${\cal L}_{kin,v}^{(b)}$, pseudoscalar initial state $B(v) = P_+
(- \gamma_5)$ and pseudoscalar final
state $\overline{D}(v') = \gamma_5P'_+$, one finds, for any current (\ref{8e})
\bea
\label{11e}
&&-2 \chi_1 (w) Tr \left [ \overline{D}(v') \Gamma B(v) \right ] \nn \\
&&= - Tr \left [ \overline{D}(v') \Gamma B(v) \right ]  \sum_{n\not= 0} {1 \over \Delta E_{1/2}^{(n)}} \xi^{(n)} (w) {<B^{(n)}(v) | O_{kin, v}^{(b)}(0)|B(v)> \over \sqrt{4m_{B^{(n)}} m_B}} 
\eea

\noi where
\beq
\label{12e}
<D(v')|\overline{h}_{v'}^c(0) \Gamma h_v^b(0)|B^{(n)}(v)>\ = - \xi^{(n)}(w) Tr \left [ \overline{D}(v') \Gamma B(v) \right ] 
\eeq

\noi that yields
\beq
\label{13e}
2 \chi_1(w) = \sum_{n\not= 0} {1 \over \Delta E_{1/2}^{(n)}}\ \xi^{(n)}(w) {<B^{(n)}(v)|O_{kin, v}^{(b)}(0)|B(v)> \over \sqrt{4m_{B^{(n)}} m_B}}\ .
\eeq

\noi Likewise, we obtain, in the case of a vector initial state $B^*(v,
\varepsilon ) = P_+ {/\hskip - 2 truemm \varepsilon}$ and a vector final state
$\overline{D}^*(v', \varepsilon ') = {/ \hskip - 2 truemm
\varepsilon}'^* P'_+$  
\beq
\label{14e}
2 \chi_1(w) = \sum_{n\not= 0} {1 \over \Delta E_{1/2}^{(n)}}\ \xi^{(n)}(w) {<B^{*(n)}(v, \varepsilon )|O_{kin, v}^{(b)}(0)|B^*(v, \varepsilon )> \over \sqrt{4m_{B^{*(n)}} m_{B^*}}}
\eeq

\noi since ${\cal L}_{kin}$ is spin-independent. In the preceding expressions the energy denominators are 
\beq
\label{15e}
\Delta E_{1/2}^{(n)} = E_{1/2}^{(n)} - E_{1/2}^{(0)} \qquad (n \not= 0) \ .
\eeq

(2) Consider ${\cal L}_{mag, v}^{(b)}$, pseudoscalar initial state $B(v) =
P_+ (- \gamma_5)$ and pseudoscalar final state $\overline{D}(v') =
\gamma_5 P'_+$. Because of parity conservation by the strong
interactions, the intermediate states $B^{(n)}$ must have the same
parity than the initial state $B$. Moreover, ${\cal L}_{mag, v}^{(b)}$ being a
scalar and producing transitions at zero recoil, the spin of $B$ and
$B^{(n)}$ must be the same. Therefore, only pseudoscalar intermediate
states $B^{(n)}(0^-)$ can contribute, only states with $j
= {1 \over 2}^-$. One finds, for any current (\ref{8e})
\bea
\label{16e}
&&4(w-1) \chi_2(w) Tr \left [ \overline{D} (v') \Gamma B(v) \right ] - 12 \chi_3 (w) Tr \left [ \overline{D} (v') \Gamma B(v) \right ] \nn \\
&&= - Tr \left [ \overline{D}(v') \Gamma B(v) \right ]  \sum_{n\not= 0} {1 \over \Delta E_{1/2}^{(n)}} \xi^{(n)} (w) {<B^{(n)}(v) | O_{mag, v}^{(b)}(0)|B(v)> \over \sqrt{4m_{B^{(n)}} m_B}} 
\eea 

\noi that gives
\beq
\label{17e}
- 4(w-1) \chi_2(w) + 12 \chi_3 (w) = \sum_{n\not= 0} {1 \over \Delta E_{1/2}^{(n)}} \xi^{(n)} (w) {<B^{(n)}(v) | O_{mag, v}^{(b)}(0)|B(v)> \over \sqrt{4m_{B^{(n)}} m_B}} \ . 
\eeq

\noi It is remarkable that this linear combination depends only on ${1
\over 2}^-$ intermediate states. We will comment on this feature
below.\par

(3) Consider ${\cal L}_{mag, v}^{(b)}$ and a vector initial state $B^*(v ,
\varepsilon ) = P_+ {/ \hskip - 2 truemm \varepsilon}$ and pseudoscalar final state
$\overline{D}(v') = \gamma_5 P'_+$. Now we will have vector
$1^-$ intermediate states, either $B^{*(n)}\left ( {1 \over 2}^-,
1^-\right )$ or $B^{*(n)}\left ( {3 \over 2}^-, 1^-\right )$. For
the latter, we have to compute the current matrix element
\beq
\label{18e}
<D(v')|J^{cb}(0)|B^{*(n)}\left ( \textstyle{{3 \over 2}^-, 1^-}\right )(v, \varepsilon )>\ = \tau_{3/2}^{(2)(n)}(w) Tr \left [ \overline{D}(v') \Gamma F_v^{\sigma} v'_{\sigma}\right ] 
\eeq

\noi where the $\left ( {3 \over 2}^-, 1^-\right )$ operator is given by
\beq
\label{19e}
F_v^{\sigma} = \sqrt{{3 \over 2}} P_+ \varepsilon_{\nu} \left [ g^{\sigma \nu} - {1 \over 3} \gamma^{\nu} \left ( \gamma^{\sigma} + v^{\sigma}\right ) \right ]
\eeq

\noi obtained from the $\left ( {3 \over 2}^+, 1^+\right )$
operator defined by Leibovich et al. (formula (2.5) of \cite{8r}),
multiplying by $(-\gamma_5)$ on the right \cite{9r}. The Isgur-Wise
functions $\tau_{3/2}^{(2)(n)}(w)$ correspond to ${1 \over 2}^- \to
{3 \over 2}^-$ transitions, the superindex (\ref{2e}) meaning the
orbital angular momentum \cite{3r} \cite{4r} \cite{9r}. As noticed by
Leibovich et al., on general grounds the IW functions
$\tau_{3/2}^{(2)(n)}(w)$ do not vanish at zero recoil.\par

One finds, for any curent (\ref{8e}), 
\bea
\label{20e}
&&<D(v')|J^{cb}(0)|B^{(n)}\left ( \textstyle{{3 \over 2}^-, 1^-}\right )(v, \varepsilon )>\ = \sqrt{{3 \over 2}} \ \tau_{3/2}^{(2)(n)}(w) (\varepsilon\cdot v') Tr \left [ \overline{D}(v') \Gamma P_+\right ]\nn \\
&&- {1 \over \sqrt{6}} (w-1) \tau_{3/2}^{(2)(n)}(w) Tr \left [ \overline{D}(v') \Gamma P_+ {/\hskip-2 truemm  \varepsilon} \right ]
\eea

\noi and finally 
\bea
\label{21e}
&&- 4 \chi_2 (w) (\varepsilon \cdot v') Tr \left [ \overline{D}(v') \Gamma P_+\right ] + 4 \chi_3 (w) Tr \left [ \overline{D}(v') \Gamma B^*(v, \varepsilon )\right ]  \nn \\
&&= - Tr \left [ \overline{D}(v') \Gamma B^*(v, \varepsilon)\right ] \sum_{n\not= 0} {1 \over \Delta E_{1/2}^{(n)}} \xi^{(n)} (w) {<B^{*(n)}(v, \varepsilon) | O_{mag, v}^{(b)}(0)|B^*(v, \varepsilon)> \over \sqrt{4m_{B^{*(n)}} m_{B^*}}} \nn \\
&&+ \left \{ \sqrt{{3 \over 2}} (\varepsilon \cdot v') Tr \left [ \overline{D}(v') \Gamma P_+\right ] - {1 \over \sqrt{6}} (w-1) Tr \left [ \overline{D}(v') \Gamma B^*(v, \varepsilon )\right ]\right \} \nn \\
&&\sum_n {1 \over \Delta E_{3/2}^{(n)}} \tau_{3/2}^{(2)(n)} (w) {<B^{*(n)}_{3/2}(v, \varepsilon) | O_{mag, v}^{(b)}(0)|B^*(v, \varepsilon)> \over \sqrt{4m_{B^{*(n)}_{3/2}} m_{B^*}}}\ . 
\eea

The energy denominators $\Delta E_{1/2}^{(n)}$ and $\Delta E_{3/2}^{(n)}$
\bea
\label{22e}
&\Delta E_{1/2}^{(n)} = E_{1/2}^{(n)} - E_{1/2}^{(0)} &\qquad (n \not= 0) \nn \\
&\Delta E_{3/2}^{(n)} = E_{3/2}^{(n)} - E_{1/2}^{(0)} &\qquad (n \geq 0) \ . 
\eea

To obtain other linearly independent relations, let us specify the
final state and the current. We make explicit the pseudoscalar $\overline{D}(v') =
\gamma_5 P'_+$ and take $\Gamma = \gamma_{\mu} \gamma_5$. This gives, from
the preceding expression,
\bea
\label{23e}
&&- 4 \chi_2 (w) (\varepsilon \cdot v') (v'_{\mu} - v_{\mu}) + 4 \chi_3 (w) \left [ (w-1) \varepsilon_{\mu} + (\varepsilon \cdot v') v_{\mu}\right ]  \nn \\
&&= - \left [ (w-1) \varepsilon_{\mu} + (\varepsilon \cdot v' ) v_{\mu}  \right ] \sum_{n\not= 0} {1 \over \Delta E_{1/2}^{(n)}} \xi^{(n)} (w) {<B^{*(n)}(v, \varepsilon) | O_{mag, v}^{(b)}(0)|B^*(v, \varepsilon)> \over \sqrt{4m_{B^{*(n)}} m_{B^*}}} \nn \\
&&+ \left \{ \sqrt{{3 \over 2}} (\varepsilon \cdot v') (v'_{\mu} - v_{\mu}) - {1 \over \sqrt{6}} (w-1) \left [ (w-1) \varepsilon_{\mu} + (\varepsilon \cdot v') v_{\mu} \right ]\right \} \nn \\
&&\sum_n {1 \over \Delta E_{3/2}^{(n)}} \tau_{3/2}^{(2)(n)} (w) {<B^{*(n)}_{3/2}(v, \varepsilon) | O_{mag, v}^{(b)}(0)|B^*(v, \varepsilon)> \over \sqrt{4m_{B^{*(n)}} m_{B^*}}}\ . 
\eea

\noi Since the two four vectors $(v'_{\mu} - v_{\mu})$ and
$[(w-1)\varepsilon_{\mu} + (\varepsilon \cdot v' ) v_{\mu}  ]$ can be
chosen to be independent, one obtains independent sum rules for
$\chi_2(w)$ and $\chi_3(w)$, namely
\beq
\label{24e}
-2 \chi_2(w) = {1 \over 2} \sqrt{{3 \over 2}} \sum_n {1 \over \Delta E_{3/2}^{(n)}} \tau_{3/2}^{(2)(n)} (w){<B^{*(n)}_{3/2}(v, \varepsilon) | O_{mag, v}^{(b)}(0)|B^*(v, \varepsilon)> \over \sqrt{4m_{B^{*(n)}_{3/2}} m_{B^*}}}
\eeq  
\bea
\label{25e}
4\chi_3 (w) &=& - \sum_{n\not= 0} {1 \over \Delta E_{1/2}^{(n)}} \xi^{(n)} (w) {<B^{*(n)}(v, \varepsilon) | O_{mag, v}^{(b)}(0)|B^*(v, \varepsilon)> \over \sqrt{4m_{B^{*(n)}} m_{B^*}}} \nn \\
&&- {w-1 \over \sqrt{6}} \sum_n {1 \over \Delta E_{3/2}^{(n)}} \tau_{3/2}^{(2)(n)} (w) {<B^{*(n)}_{3/2}(v, \varepsilon) | O_{mag, v}^{(b)}(0)|B^*(v, \varepsilon)> \over \sqrt{4m_{B^{*(n)}_{3/2}} m_{B^*}}}
\eea

As a final remark on this Section on the derivation of the sum
rules, let us point out that if, instead of (\ref{5e}) that involves
${\cal L}^{(b)}$ we start from (\ref{6e}) with ${\cal L}^{(c)}$, we
obtain the same SR as above, with the replacement $b \to c$ in the
operators and in the states. The reason is that the IW functions and
energy denominators are flavor-independent in the heavy quark limit. 

\section{Summary and comments on the Lagrangian sum rules.}
\hspace*{\parindent} To summarize, making explicit the $c$ flavor, we
have obtained the sum rules 
\bea
\label{26e}
&&\chi_1(w) = {1 \over 2} \sum_{n\not= 0} {1 \over \Delta E_{1/2}^{(n)}} \xi^{(n)} (w) {<D^{(n)}(v) | O_{kin, v}^{(c)}(0)|D(v)> \over \sqrt{4m_{D^{(n)}} m_{D}}} \nn \\
&&\qquad = {1 \over 2} \sum_{n\not= 0} {1 \over \Delta E_{1/2}^{(n)}} \xi^{(n)} (w) {<D^{*(n)}(v, \varepsilon) | O_{kin, v}^{(c)}(0)|D^*(v, \varepsilon)> \over \sqrt{4m_{D^{*(n)}} m_{D^*}}} \\ 
&& \nn \\
\label{27e}
&&\chi_2(w) = - {3 \over 4 \sqrt{6}} \sum_n {1 \over \Delta E_{3/2}^{(n)}} \tau_{3/2}^{(2)(n)} (w) {<D^{*(n)}_{3/2}(v, \varepsilon) | O_{mag, v}^{(c)}(0)|D^*(v, \varepsilon)> \over \sqrt{4m_{D^{*(n)}_{3/2}} m_{D^*}}} \\
&& \nn \\
&&\chi_3(w) = - {1 \over 4} \sum_{n\not= 0} {1 \over \Delta E_{1/2}^{(n)}} \xi^{(n)} (w) {<D^{*(n)}(v, \varepsilon) | O_{mag, v}^{(c)}(0)|D^*(v, \varepsilon)> \over \sqrt{4m_{D^{*(n)}} m_{D^*}}} \nn \\
&&\qquad - \ {w-1 \over 4\sqrt{6}} \sum_{n} {1 \over \Delta E_{3/2}^{(n)}} \tau_{3/2}^{(2)(n)} (w) {<D^{*(n)}_{3/2}(v, \varepsilon) | O_{mag, v}^{(c)}(0)|D^*(v, \varepsilon)> \over \sqrt{4m_{D^{*(n)}_{3/2}} m_{D^*}}} 
\label{28e}
\eea

There are a number of striking features in relations
(\ref{26e})-(\ref{28e}).\par

(i) One should notice that {\it elastic subleading form factors of the
Lagrangian type} are given in terms of {\it leading IW functions},
namely $\xi^{(n)}(w)$ and $\tau_{3/2}^{(2)(n)}(w)$, and {\it
subleading} form factors {\it at zero recoil}. \par

(ii) $\chi_1 (w)$ is given in terms of matrix elements of ${\cal
L}_{kin}$, as expected from the definitions (\ref{5e})-(\ref{6e}) and
involve transitions ${1\over 2}^- \to {1 \over 2}^-$. \par

(iii) The {\it elastic subleading magnetic form factors} $\chi_2(w)$ and
$\chi_3(w)$ involve $D^*(1^-) \to D^{*(n)} (1^-)$ transitions ${1 \over 2}^- \to {1 \over 2}^-$ and ${1 \over 2}^- \to {3 \over 2}^-$.\par

(iv) $\chi_1(w)$ and $\chi_3(w)$ satisfy, as they should, Luke theorem
\cite{10r},
\beq
\label{29e}
\chi_1(1) = \chi_3(1) = 0
\eeq

\noi because the ${1 \over 2}^- \to {1 \over 2}^-$ IW functions at zero recoil satisfy
\beq
\label{30e}
\xi^{(n)} (1) = \delta_{n,0}
\eeq

(v) There is a linear combination of $\chi_2(w)$ and $\chi_3(w)$ that
gets only contributions from ${1 \over 2}^- \to {1 \over 2}^-$
transitions, namely
\beq
\label{31e}
-4(w-1)\chi_2(w) + 12 \chi_3(w) = - 3 \sum_{n\not= 0} {1 \over \Delta E_{1/2}^{(n)}} \xi^{(n)} (w) {<D^{*(n)}(v, \varepsilon) | O_{mag, v}^{(c)}(0)|D^*(v, \varepsilon)> \over \sqrt{4m_{D^{*(n)}} m_{D^*}}}  
\eeq

\noi where the factor $-3$ is in consistency with (\ref{17e}), shifting
from vector to pseudoscalar mesons. \par

This latter relation and (\ref{26e}) imply that the combination 
\beq
\label{32e}
L_1(w) = 2 \chi_1(w) - 4(w-1) \chi_2(w) + 12 \chi_3(w)
\eeq

\noi gets only contributions from ${1 \over 2}^- \to {1 \over 2}^-$
transitions. We will give an alternative demonstration of this feature using the OPE in Appendix A.

\section{The OPE sum rule for h$_{\bf A_1}$(1).} \hspace*{\parindent} 
It is well-known that the determination of $|V_{cb}|$ from the $\overline{B} \to
D^*\ell \nu$ differential rate at zero recoil depends on the value of
$h_{A_1}(1)$. \par

The interesting point is that precisely the subleading matrix elements
of $O_{kin}$ and $O_{mag}$ at zero recoil, that enter in the SR
(\ref{26e})-(\ref{28e}), are related to the quantity $|h_{A_1}(1)|$, as we will see now.

The following SR follows from the OPE \cite{11r} \cite{8r},
\bea
\label{33e}
&&|h_{A_1}(1)|^2 + \sum_n {|<D^{*(n)}\left ( {1 \over 2}^-, {3 \over 2}^-\right ) (v, \varepsilon )| \vec{A}|B(v)>|^2 \over 4m_{D^{*(n)}} m_B}\nn \\
&&= \eta_A^2 - {\mu_G^2 \over 3 m_c^2} - {\mu_{\pi}^2 - \mu_G^2 \over 4} \left ( {1 \over m_c^2} + {1 \over m_b^2} + {2 \over 3m_cm_b}\right )
\eea

\noi where $D^{*(n)}$ are $1^-$ excited states, and 
\bea
\label{34e}
&&\mu_{\pi}^2 = {1 \over 2m_B} \ <B(v)|\overline{h}_v^{(b)}(iD)^2 h_v^{(b)}|B(v)> \nn \\
&&\mu_{G}^2 = {1 \over 2m_B} \ <B(v)|\overline{h}_v^{(b)}{g_s \over 2} \sigma_{\alpha\beta}G^{\alpha\beta} h_v^{(b)}|B(v)> \nn \\
&& \quad = - {3 \over 2m_B} \ <B^*(v, \varepsilon ) |\overline{h}_v^{(b)}{g_s \over 2} \sigma_{\alpha\beta} G^{\alpha\beta} h_v^{(b)}|B^*(v, \varepsilon )> 
\eea

\noi In relation (\ref{33e}) one assumes the states at rest $v = (1,
{\bf 0})$ and the axial current is space-like, orthogonal to $v$. The relation of (\ref{34e}) with the other common notation \cite{18r} \cite{8r} is $\mu_{\pi}^2 = - \lambda_1$ and $\mu_G^2 = 3 \lambda_2$.\par

In the l.h.s. of relation (\ref{33e}), 
\beq
\label{35e}
h_{A_1}(1) = \eta_{A_1} + \delta_{1/m^2}^{(A_1)}
\eeq

\noi ($\eta_{A_1} = 1 +$ radiative corrections) because there are no first order $1/m_Q$ corrections due to Luke theorem \cite{10r}. The sum over the
squared matrix elements of $B \to D^{*(n)}(1^-)$ transitions contains
two types of possible contributions, corresponding to $D^{*(n)}\left (
{1 \over 2}^-, 1^-\right )$ $(n \not= 0)$, and $D^{*(n)}\left ( {3
\over 2}^-, 1^-\right )$ $(n \geq 0)$. The r.h.s. of (\ref{33e})
exhibits the OPE at the desired order. From the decomposition between
radiative corrections and $1/m_Q^2$ corrections (\ref{35e}) one gets,
from (\ref{33e}), neglecting higher order terms,
\bea
\label{36e}
&&- \delta_{1/m^2}^{(A_1)} = {\mu_G^2 \over 6 m_c^2} + {\mu_{\pi}^2 - \mu_G^2 \over 8} \left ( {1 \over m_c^2} + {1 \over m_b^2} + {2 \over 3m_cm_b}\right )  \nn \\
&&+ {1 \over 2} \sum_n {|<D^{*(n)}\left ( {1 \over 2}^-, {3 \over 2}^-\right ) (v, \varepsilon )| \vec{A}|B(v)>|^2 \over 4m_{D^{*(n)}} m_B} \ .
\eea

\noi The correction $\delta_{1/m^2}^{(A_1)}$ is therefore negative, both terms
being of the same sign.\par

The matrix elements $<D^{*(n)}\left ( {1 \over 2}^-, {3 \over
2}^-\right ) (v, \varepsilon )| \vec{A}|B>$ have been expressed in
terms of the matrix elements $<D^{*(n)}\left ( {1 \over 2}^-\right )
(v, \varepsilon )|O_{kin, v}^{(c)}(0)|D^*(v, \varepsilon )>$ and
$<D^{*(n)}\left ( {1 \over 2}^-, {3 \over 2}^-\right )$\break \noindent $(v, \varepsilon
)|O_{mag, v}^{(c)}(0)|D^*(v, \varepsilon )>$ by Leibovich et al.
\cite{8r}, within the same normalization convention used in the
preceding sections,
\bea
\label{37e}
&&{<D^{*(n)}\left ( {1 \over 2}^-\right ) (v, \varepsilon )| \vec{A}|B(v)> \over \sqrt{4m_{D^{*(n)}} m_B}} \nn \\
&&= - {\vec{\varepsilon} \over \Delta E_{1/2}^{(n)}} \left [ \left ( {1 \over 2m_c} + {3 \over 2m_b}\right ) {<D^{*(n)}\left ( {1 \over 2}^-\right ) (v, \varepsilon) | O_{mag, v}^{(c)}(0)|D^*(v, \varepsilon)> \over \sqrt{4m_{D^{*(n)}} m_{D^*}}} \right .\nn \\
&&\left . + \left ( {1 \over 2m_c} - {1 \over 2m_b}\right ) {<D^{*(n)}\left ( {1 \over 2}^-\right )(v, \varepsilon) | O_{kin, v}^{(c)}(0)|D^*(v, \varepsilon)> \over \sqrt{4m_{D^{*(n)}} m_{D^*}}} \right ] 
\eea
\bea
\label{38e}
&&{<D^{*(n)}\left ( {3 \over 2}^-\right ) (v, \varepsilon )| \vec{A}|B(v)> \over \sqrt{4m_{D^{*(n)}} m_B}}  \nn \\ 
&&= - {\vec{\varepsilon} \over \Delta E_{3/2}^{(n)}} {1 \over 2m_c}  {<D^{*(n)}\left ( {3 \over 2}^-\right ) (v, \varepsilon) | O_{mag, v}^{(c)}(0)|D^*(v, \varepsilon)> \over \sqrt{4m_{D^{*(n)}_{3/2}} m_{D^*}}}\ . 
\eea

\noi Therefore $-\delta_{1/m^2}^{(A_1)}$ (\ref{36e}) can be written as 
\bea
\label{39e}
&&- \delta_{1/m^2}^{(A_1)} = {\mu_G^2 \over 6 m_c^2} + {1 \over 8} \left ( {1 \over m_c^2} + {1 \over m_b^2} + {2 \over 3m_cm_b}\right ) \left ( \mu_{\pi}^2 - \mu_G^2\right ) \nn \\
&&+ {1 \over 2} \sum_n \left [ \left ( {1 \over 2m_c} + {3 \over 2m_b}\right ) {1 \over \Delta E_{1/2}^{(n)}}\ {|<D^{*(n)}\left ( {1 \over 2}^-\right ) (v, \varepsilon )| O_{mag, v}^{(c)}(0)|D^*(v, \varepsilon)> \over \sqrt{4m_{D^{*(n)}} m_{D^*}}} \right . \nn \\
&&\left . + \left ( {1 \over 2m_c} - {1 \over 2m_b}\right ) {1 \over \Delta E_{1/2}^{(n)}}\ {<D^{*(n)}\left ( {1 \over 2}^-\right ) (v, \varepsilon )| O_{kin, v}^{(c)}(0)|D^*(v, \varepsilon)> \over \sqrt{4m_{D^{*(n)}} m_{D^*}}} \right ]^2 \nn \\
&&+ {1 \over 2} \sum_n \left [  {1 \over 2m_c} \ {1 \over \Delta E_{3/2}^{(n)}}\ {<D^{*(n)}\left ( {3 \over 2}^-\right ) (v, \varepsilon )| O_{mag, v}^{(c)}(0)|D^*(v, \varepsilon)> \over \sqrt{4m_{D^{*(n)}_{3/2}} m_{D^*}}} \right ]^2\ .
\eea

The important point to emphasize here is that the matrix elements\break\noindent $<D^{*(n)}\left ( {1 \over 2}^-\right ) (v,
\varepsilon )| O_{kin, v}^{(c)}(0)|D^*(v, \varepsilon)>$ and
$<D^{*(n)}\left ( {1 \over 2}^-, {3 \over 2}^-\right )(v, \varepsilon
)| O_{mag, v}^{(c)}(0)|D^*(v, \varepsilon)>$ are precisely the same ones
that enter in the SR (\ref{26e})-(\ref{28e}). This allows to obtain an
interesting lower bound on $-\delta_{1/m^2}^{(A_1)}$.

\section{A lower bound on the inelastic contribution to the $-\delta_{\bf 1/m^2}^{\bf (A_1)}$ correction of the B $\to$ D$^{\bf *}$ axial form factor at zero recoil.} \hspace*{\parindent} 
We take now the relevant linear combinations of the matrix elements suggested by the r.h.s. of (\ref{39e}), and use (\ref{26e}), (\ref{27e}) and (\ref{31e}),
\bea
\label{40e}
&&\sum_{n\not= 0} {1 \over \Delta E_{1/2}^{(n)}} \xi^{(n)}(w) \left \{ \left ( {1 \over 2m_c} - {1 \over 2m_b}\right ) {<D^{*(n)}\left ( {1 \over 2}^-\right ) (v, \varepsilon) | O_{kin, v}^{(c)}(0)|D^*(v, \varepsilon)> \over \sqrt{4m_{D^{*(n)}} m_{D^*}}}\right . \nn \\
&&\left . + \left ( {1 \over 2m_c} + {3 \over 2m_b}\right ) {<D^{*(n)}\left ( {1 \over 2}^-\right ) (v, \varepsilon) | O_{mag, v}^{(c)}(0)|D^*(v, \varepsilon)> \over \sqrt{4m_{D^{*(n)}} m_{D^*}}}\right \} \nn \\
&&= \left ( {1 \over 2m_c} - {1 \over 2m_b}\right ) 2\chi_1(w) - {1 \over 3} \left ( {1 \over 2m_c} + {3 \over 2m_b}\right ) \left [ - 4(w-1) \chi_2 (w) + 12 \chi_3 (w) \right ] \nn \\
\eea
\bea
\label{41e}
&&\sum_n {1 \over \Delta E_{3/2}^{(n)}} \tau_{3/2}^{(2)(n)} (w){<D^{*(n)}\left ( {3 \over 2}^-\right )(v, \varepsilon) | O_{mag, v}^{(c)}(0)|D^*(v, \varepsilon)> \over \sqrt{4m_{D^{*(n)}_{3/2}} m_{D^*}}}\nn \\
&&= - {1 \over 2m_c} \ {4 \sqrt{6} \over 3} \ \chi_2(w)\ .
\eea

\noi Using now Schwarz inequality
\beq
\label{42e}
\left | \sum_n A_n B_n\right | \leq \sqrt{\left ( \sum_n |A_n|^2\right ) \left ( \sum_n |B_n|^2\right )}
\eeq
 
\noi one finds
$$\sum_{n\not=0} \left [ \xi^{(n)} (w)\right ]^2 \sum_{n\not= 0}
\left \{ {1 \over \Delta E_{1/2}^{(n)}} \left [ \left ( {1 \over 2m_c} - {1 \over 2m_b}\right ) {<D^{*(n)}\left ( {1 \over 2}^-\right ) (v, \varepsilon )| O_{kin, v}^{(c)}(0)|D^*(v, \varepsilon)> \over \sqrt{4m_{D^{*(n)}} m_{D^*}}} \right . \right .$$
\bea
\label{43e}
&&\left . \left . + \left ( {1 \over 2m_c} + {3 \over 2m_b}\right ) {<D^{*(n)}\left ( {1 \over 2}^-\right ) (v, \varepsilon )| O_{mag, v}^{(c)}(0)|D^*(v, \varepsilon)> \over \sqrt{4m_{D^{*(n)}} m_{D^*}}} \right ] \right \}^2 \nn \\
&& \geq 4\left \{ \left ( {1 \over 2m_c} - {1 \over 2m_b}\right )\chi_1(w) - {1 \over 3} \left ( {1 \over 2m_c} + {3 \over 2m_b}\right ) \left [ - 2(w-1) \chi_2 (w)+ 6 \chi_3 (w) \right ] \right \}^2 \nn \\
 \eea
\bea
\label{44e}
&&\sum_n \left [ \tau_{3/2}^{(2)(n)}(w)\right ]^2 \sum_n \left \{ {1 \over \Delta E_{3/2}^{(n)}} \left [ {1 \over 2m_c} \ {<D^{*(n)}\left ( {3 \over 2}^-\right ) (v, \varepsilon )| O_{mag, v}^{(c)}(0)|D^*(v, \varepsilon)> \over \sqrt{4m_{D^{*(n)}_{3/2}} m_{D^*}}}\right ] \right \}^2 \nn \\
&&\geq {32 \over 3} \left [ {1 \over 2m_c} \ \chi_2(w) \right ]^2\ .
\eea
\vskip 3 truemm

These two last equations imply, from (\ref{39e}), the inequality 
\bea
\label{45e}
&&- \delta_{1/m^2}^{(A_1)} \geq {\mu_G^2 \over 6 m_c^2} + {\mu_{\pi}^2 - \mu_G^2 \over 8} \left ( {1 \over m_c^2} + {1 \over m_b^2} + {2 \over 3m_cm_b}\right )  \nn \\
&&+ \ 2 {\left \{ \left ( {1 \over 2m_c} - {1 \over 2m_b}\right ) \chi_1(w) - {1 \over 3} \left ( {1 \over 2m_c} + {3 \over 2m_b}\right ) \left [-2(w-1)\chi_2(w) + 6\chi_3 (w) \right ] \right \}^2 \over \sum\limits_{n\not= 0} \left [ \xi^{(n)}(w)\right ]^2} \nn \\
&& + {16 \over 3} {\left [ {1 \over 2m_c}  \chi_2(w)\right ]^2 \over \sum\limits_n \left [ \tau_{3/2}^{(2)(n)}(w)\right ]^2}\ .
\eea

This inequality on $-\delta_{1/m^2}^{(A_1)}$ involves on the r.h.s. {\it elastic subleading} functions $\chi_i(w)$ $(i = 1,2,3)$ in the
numerator and sums over {\it inelastic leading IW functions}
$\sum\limits_{n\not= 0} [\xi^{(n)}(w)]^2$ and $\sum\limits_n
[\tau_{3/2}^{(2)(n)}(w)]^2$ in the denominator. We must emphasize that
this inequality is valid for all values of $w$ and constitutes a
rigorous constraint between these functions and the correction
$-\delta_{1/m^2}^{(A_1)}$. Let us point out that, near $w=1$, since 
\beq
\label{46e}
\xi^{(n)}(w) \sim (w-1) \qquad\qquad (n \not= 0)
\eeq

\noi and, due to Luke theorem
\beq
\label{47e}
\chi_1(w), \ \chi_3(w) \sim (w-1)
\eeq

\noi the second term on the r.h.s. of (\ref{45e}) is a constant in the
limit $w \to 1$. \par

On the other hand, since $\chi_2(w)$ is not protected by Luke theorem,
\beq
\label{48e}
\chi_2(1) \not= 0
\eeq

\noi and in general, as pointed out by Leibovich et al. \cite{8r}
\beq
\label{49e}
\tau_{3/2}^{(2)}(1) \not= 0
\eeq

\noi the last term in the r.h.s. of (\ref{45e}) is also a constant for
$w = 1$. \par

The inequality (\ref{45e}) is valid for all values of $w$, and in
particular it holds in the $w \to 1$ limit. Let us consider this limit,
that gives
\bea
\label{50e}
&&- \delta_{1/m^2}^{(A_1)} \geq {\mu_G^2 \over 6 m_c^2} + {\mu_{\pi}^2 - \mu_G^2 \over 8} \left ( {1 \over m_c^2} + {1 \over m_b^2} + {2 \over 3m_cm_b}\right ) \nn \\
&&+2 {\left \{ \left ( {1 \over 2m_c} - {1 \over 2m_b}\right ) \chi '_1(1) - {1 \over 3} \left ( {1 \over 2m_c} + {3 \over 2m_b}\right ) \left [-2\chi_2(1) + 6\chi '_3 (1) \right ] \right \}^2 \over \sum\limits_{n\not= 0} \left [ \xi ^{(n)'}(1)\right ]^2} \nn \\
&& + {16 \over 3} {\left [ {1 \over 2m_c}  \chi_2(1)\right ]^2 \over \sum\limits_n \left [ \tau_{3/2}^{(2)}(1)\right ]^2}\ .
\eea

\noi On the other hand, using the OPE in the heavy quark limit, we have demonstrated the following sum rules \cite{5r}
\beq
\label{51e}
\sum_n \left [ \tau_{3/2}^{(2)}(1)\right ]^2 = {4 \over 5} \sigma^2 - \rho^2 
\eeq
\beq
\label{52e}
\sum_{n\not= 0} \left [ \xi^{(n)'}(1)\right ]^2 = {5 \over 3} \sigma^2 - {4 \over 3 }\rho^2 - (\rho^2)^2 
\eeq

\noi where $\rho^2$ and $\sigma^2$ are the slope and the curvature of the elastic Isgur-Wise function $\xi (w)$,
\beq
\label{53e}
\xi (w) = 1 - \rho^2 (w-1) + {\sigma^2 \over 2} (w-1)^2 + \cdots
\eeq

\noi The positivity of the l.h.s. of (\ref{51e}), (\ref{52e}) yield respectively the lower bounds on the curvature obtained in \cite{4r} \cite{5r},
\beq
\label{54new}
\sigma^2 \geq {5 \over 4} \rho^2 
\eeq

\beq
\label{54e}
\sigma^2 \geq {1 \over 5} \left [ 4 \rho^2 + 3 (\rho^2)^2\right ] \ .
\eeq

On the other hand, Uraltsev \cite{1r} plus Bjorken \cite{2r} SR imply
\beq
\label{55new}
\rho^2 \geq {3 \over 4}
\eeq

\noi giving, from both (\ref{54new}), (\ref{54e}), the absolute bound for the curvature
\beq
\label{56new}
\sigma^2 \geq {15 \over 16} \ . 
\eeq

\noi Relations (\ref{50e})-(\ref{52e}) give finally the bound 
\bea
\label{55e}
&&- \delta_{1/m^2}^{(A_1)} \geq {\mu_G^2 \over 6 m_c^2} + {\mu_{\pi}^2 - \mu_G^2 \over 8} \left ( {1 \over m_c^2} + {1 \over m_b^2} + {2 \over 3m_cm_b}\right )  \nn \\
&&+ {2 \over 3[5 \sigma^2 - 4 \rho^2 - 3(\rho^2)^2]} \left \{ \left ( {1 \over 2m_c} - {1 \over 2m_b}\right ) 3\chi '_1(1) -  \left ( {1 \over 2m_c} + {3 \over 2m_b}\right ) \left [-2\chi_2(1) + 6\chi '_3 (1) \right ] \right \}^2\nn \\
&& + {80 \over 3(4\sigma^2 - 5\rho^2)} \left [ {1 \over 2m_c} \chi_2(1)\right ]^2  \ .
\eea

We briefly discuss in Appendix B the radiative corrections to relations
(\ref{51e}) and (\ref{52e}), computed in \cite{6r}, and their impact on
the bound (\ref{55e}).

\section{Lower bound on  the 1/m$_{\bf Q}^{\bf 2}$ corrections to h$_{\bf +}$(1) and h$_{\bf 1}$(1).} \hspace*{\parindent} 
Of theoretical interest are also the quantities at zero recoil
$\ell_1(1)$, $\ell_2(1)$, that would correspond to the wave function
overlaps in the non-relativistic quark model \cite{18r}. Using the
notation of Falk and Neubert \cite{18r}, these quantities are related to
the matrix elements of the vector current at zero recoil,

\bea
\label{65e}
&&{<D(v)|V_{\mu}|B(v)> \over \sqrt{4m_Bm_D}} = v_{\mu} \ h_+(1) \nn \\
&&{<D^*(v, \varepsilon)|V_{\mu}|B^*(v, \varepsilon)> \over \sqrt{4m_{B^*}m_{D^*}}} = v_{\mu} \ h_1(1) 
\eea

\noi where
\bea
\label{65new}
&&h_+(1) = 1 + \delta_{1/m^2}^{(+)} + \cdots \nn \\
&&h_1(1) = 1 + \delta_{1/m^2}^{(1)} + \cdots
\eea
\noi with
\bea
\label{66new}
&&\delta_{1/m^2}^{(+)} = \left ( {1 \over 2m_c} - {1 \over 2m_b}\right )^2 \ell_1(1) \nn \\
&&\delta_{1/m^2}^{(1)} = \left ( {1 \over 2m_c} - {1 \over 2m_b}\right )^2 \ell_2(1) \ .
\eea
\vskip 3 truemm

\noi On the other hand, using also the notations of \cite{18r}, $\delta_{1/m^2}^{(A_1)}$ is given by the expression
\beq
\label{66e}
\delta_{1/m^2}^{(A_1)} =  \left ( {1 \over 2m_c} - {1 \over 2m_b}\right ) \left [ {1 \over 2m_c} \ell_2(1) - {1 \over 2m_b} \ell_1(1) \right ] + {1 \over 4m_cm_b} \ \Delta
\eeq

\noi where
\beq
\label{67e}
\Delta = \ell_1 (1) + \ell_2(1) + m_2(1) + m_9(1) \ .
\eeq

\noi We observe that 
\bea
\label{68e}
& \delta_{1/m^2}^{(A_1)} \to  \displaystyle{{1 \over 4m_b^2}} \ell_1(1) &\qquad \hbox{for} \quad m_c \to \infty \nn \\
& \delta_{1/m^2}^{(A_1)} \to  \displaystyle{{1 \over 4m_c^2}} \ell_2(1) &\qquad \hbox{for} \quad m_b \to \infty \ .
\eea

\noi Therefore, since the lower bound (\ref{55e}) is valid for any
value of $m_c$ and $m_b$, we can obtain lower bounds on $-\ell_1(1)$
and $-\ell_2(1)$ by taking the limits (\ref{68e}). We
find, in this way,
\beq
\label{69e}
- \ell_1(1) \geq {\mu_{\pi}^2 - \mu_G^2 \over 2}  + {6 \over 5 \sigma^2 - 4\rho^2 - 3(\rho^2)^2} \left [ - \chi '_1 (1) + 2 \chi_2(1) - 6 \chi '_3(1) \right ]^2
\eeq

$$- \ell_2(1) \geq   {3\mu_{\pi}^2 + \mu_G^2 \over 6}  + {2 \over 3[5 \sigma^2 - 4\rho^2 - 3(\rho^2)^2]} \left [ 3 \chi '_1 (1) + 2 \chi_2(1) - 6 \chi '_3(1) \right ]^2$$
\beq
\label{70e}
+ {80 \over 3 (4 \sigma^2 - 5 \rho^2)} \left [ \chi_2 (1) \right ]^2
\eeq

\noi and from (\ref{66new}) we obtain lower bounds on $-\delta_{1/m^2}^{(+)}$ and $-\delta_{1/m^2}^{(1)}$.

\section{General considerations on the bounds.}\hspace*{\parindent}
We have obtained lower bounds on the $-\delta_{1/m^2}$ corrections to
some form factors, namely $h_{A_1}(w)$, $h_+(w)$ and $h_1(w)$, {\it that are
protected by Luke theorem}. It is worth to summarize their expressions
at zero recoil~:

\bea
\label{67}
&&-\delta_{1/m^2}^{(A_1)} \geq {\mu_G^2 \over 6m_c^2} + {\mu_{\pi}^2 - \mu_G^2 \over 8} \left ( {1 \over m_c^2} + {1 \over m_b^2} + {2 \over 3m_cm_b}\right ) \nn \\
&&+ {2 \over 3[5\sigma^2 - 4 \rho^2 - 3(\rho^2)^2]} \left \{ \left ( {1 \over 2m_c} - {1 \over 2m_b}\right ) 3 \chi'_1(1) \right .\nn \\
&&\left . - \left ( {1 \over 2m_c} + {3 \over 2m_b}\right ) \left [ - 2 \chi_2(1) + 6 \chi '_3(1) \right ] \right \}^2 + {80 \over 4 \sigma^2 - 5 \rho^2} \left [ {1 \over 2m_c} \ \chi_2(1) \right ]^2\\
&&\nn \\
\label{68}
&&-\delta_{1/m^2}^{(+)} \geq \left ( {1 \over 2m_c} - {1 \over 2m_b} \right )^2\left \{ { \mu_{\pi}^2 - \mu_G^2\over 2}\right . \nn \\
&&\left.  + {6 \over 5\sigma^2 - 4 \rho^2 - 3(\rho^2)^2} \left [ - \chi'_1(1) + 2\chi_2(1) - 6 \chi '_3(1) \right ]^2 \right \} \\
&&\nn \\
&&-\delta_{1/m^2}^{(1)}  \geq \left ( {1 \over 2m_c} - {1 \over
2m_b} \right )^2 \left \{ {3\mu_{\pi}^2 + \mu_G^2 \over 6} \right . \nn \\
&&\left . + {2 \over 3[5\sigma^2 - 4 \rho^2 - 3(\rho^2)^2]} \left [ 3
\chi'_1(1) + 2\chi_2(1) - 6 \chi '_3(1) \right ]^2 + {80 \over 3(4\sigma^2 - 5 \rho^2)} [\chi_2(1)]^2 \right \} \ .\nn \\
\label{69}
\eea

A number of remarks are worth to be made here : \par

(i) The bounds contain an OPE piece, dependent on $\mu_{\pi}^2$ and
$\mu_G^2$, and a piece that bounds the inelastic contributions, given
in terms of the $1/m_Q$ elastic quantities $\chi '_1(1)$, $\chi_2(1)$,
$\chi '_3(1)$ and on the slope $\rho^2$ and curvature $\sigma^2$ of the
elastic IW function $\xi (w)$.\par

(ii) Taking roughly constant values for $\chi '_1(1)$, $\chi_2(1)$,
$\chi '_3(1)$, as suggesed by the QCD Sum Rules calculations (QCDSR)
\cite{12r} \cite{13r} \cite{14r}, the bounds for the inelastic
contributions diverge in the limit $\rho^2 \to {3 \over 4}$, $\sigma^2 \to {15 \over 16}$, according to (\ref{56new}). This
feature does not seem to us physical.\par 

(iii) {\it Therefore, one should expect that $\chi '_1(1)$, $\chi_2(1)$ and $\chi '_3(1)$ vanish also in this limit}. We give a demonstration of this interesting feature in the next section.\par
 
(iv) Thus, the limit $\rho^2 \to {3 \over 4}$, $\sigma^2 \to {15 \over 16}$ seems related to the behaviour of $\chi_i(w)$ ($i = 1, 2, 3$) near zero recoil. \par

(v) The feature (iii) does not appear explicitly  in the QDCSR approach, where one gets roughly $\rho_{ren}^2 \cong
0.7$, and where there is no dependence on $\rho^2$ of the functions
$\chi_i(w)$ ($i = 1,2,3$).\par

(vi) In the nonrelativistic quark model the parameters $\ell_1(1)$ and $\ell_2(1)$ correspond to the overlap of the wave functions at zero recoil \cite{18r}. The formulas (\ref{69e})
and (\ref{70e}) give a model-independent, rigorous bound for these
quantities.

\section{Behaviour of the subleading functions $\chi_{\bf i}$(w) \break\noindent (i = 1,2,3) in the limit $\rho^2 \to {\bf 3 \over 4}$, $\sigma^2 \to {\bf 15 \over 16}$.} \hspace*{\parindent} 
In this Section we demonstrate that indeed $\chi '_1(1)$, $\chi_2(1)$
and $\chi '_3(1)$ vanish in the limit $\rho^2 \to {3 \over 4}$, $\sigma^2 \to {15 \over 16}$. Let us
rewrite the relations (\ref{26e}), (\ref{27e}) and (\ref{31e}) in terms
of pseudoscalar matrix elements  
\beq
\label{equation70}
\chi_1(w) = {1 \over 2} \sum_{n\not= 0} {1 \over \Delta E_{1/2}^{(n)}} \xi^{(n)}(w) {<D^{(n)} (v) |O_{kin}^{(c)} (0)|D(v)> \over \sqrt{4m_{D^{(n)}}m_D}}
\eeq

\beq
\label{equation71}
\chi_2(w) = {1 \over 4\sqrt{6}} \sum_{n} {1 \over \Delta E_{3/2}^{(n)}} \tau_{3/2}^{(2)(n)}(w) {<D^{*(n)}_{3/2} (v, \varepsilon) |O_{mag}^{(c)} (0)|D^*(v, \varepsilon )> \over \sqrt{4m_{D^{*(n)}_{3/2}}m_{D^*}}}
\eeq

\beq
\label{equation72}
-4(w-1)\chi_2(w) + 12 \chi_3(w) = \sum_{n\not= 0} {1 \over \Delta E_{1/2}^{(n)}} \xi^{(n)}(w) {<D^{(n)} (v) |O_{mag}^{(c)} (0)|D(v)> \over \sqrt{4m_{D^{(n)}}m_D}}
\eeq

\noi At zero recoil $w \to 1$ we have
\beq
\label{equation73}
\chi '_1(1) = {1 \over 2} \sum_{n\not= 0} {1 \over \Delta E_{1/2}^{(n)}} \xi^{(n)'}(1) {<D^{(n)} (v) |O_{kin}^{(c)} (0)|D(v)> \over \sqrt{4m_{D^{(n)}}m_D}}
\eeq

\beq
\label{equation74}
\chi_2(1) = {1 \over 4\sqrt{6}} \sum_{n} {1 \over \Delta E_{3/2}^{(n)}} \tau_{3/2}^{(2)(n)}(1) {<D^{*(n)}_{3/2} (v, \varepsilon ) |O_{mag}^{(c)} (0)|D^*(v, \varepsilon )> \over \sqrt{4m_{D^{*(n)}_{3/2}}m_{D^*}}}
\eeq

\beq
\label{equation75}
-4 \chi_2(1) + 12\chi '_3(1) = \sum_{n\not= 0} {1 \over \Delta E_{1/2}^{(n)}} \xi^{(n)'}(1) {<D^{(n)} (v) |O_{mag}^{(c)} (0)|D(v)> \over \sqrt{4m_{D^{(n)}}m_D}}
\eeq

\noi Using again Schwarz inequality as in Section 5, we obtain 
\beq
\label{equation76}
[\chi '_1(1)]^2 \leq  {1 \over 4} \sum_{n\not= 0} \left [\xi^{(n)'}(1)\right ]^2\sum_{n\not= 0} \left [ {1 \over \Delta E_{1/2}^{(n)}}  {<D^{(n)} (v) |O_{kin}^{(c)} (0)|D(v)> \over \sqrt{4m_{D^{(n)}}m_D}}\right ]^2
\eeq

\beq
\label{equation77}
[\chi_2(1)]^2 \leq  {1 \over 96} \sum_{n} \left [ \tau_{3/2}^{(2)(n)}(1)\right ]^2\sum_{n} \left [ {1 \over \Delta E_{3/2}^{(n)}}  {<D^{*(n)}_{3/2} (v, \varepsilon) |O_{mag}^{(c)} (0)|D^*(v, \varepsilon)> \over \sqrt{4m_{D^{(n)}}m_{D^*}}}\right ]^2
\eeq

\beq
\label{equation78}
\left [ -4 \chi_2(1) + 12\chi '_3(1)\right  ]^2 \leq  \sum_{n\not= 0}  \left [ \xi^{(n)'}(1) \right ]^2 \sum_{n\not= 0} \left [ {1 \over \Delta E_{1/2}^{(n)}}{<D^{(n)} (v) |O_{mag}^{(c)} (0)|D(v)> \over \sqrt{4m_{D^{(n)}}m_D}}\right ]^2
\eeq

\noi and from relations (\ref{51e}) and (\ref{52e}) we obtain
\beq
\label{equation79}
[\chi '_1(1)]^2 \leq  {1 \over 12}  \left [5 \sigma^2 - 4 \rho^2 - 3(\rho^2)^2\right ] \sum_{n\not= 0} \left [ {1 \over \Delta E_{1/2}^{(n)}}  {<D^{(n)} (v) |O_{kin}^{(c)} (0)|D(v)> \over \sqrt{4m_{D^{(n)}}m_D}}\right ]^2
\eeq

\beq
\label{equation80}
[\chi_2(1)]^2 \leq {1 \over 480}  \left (4 \sigma^2 - 5 \rho^2) \right )\sum_{n} \left [ {1 \over \Delta E_{3/2}^{(n)}}  {<D^{*(n)}_{3/2} (v, \varepsilon ) |O_{mag}^{(c)} (0)|D^*(v, \varepsilon )> \over \sqrt{4m_{D^{*(n)}_{3/2}}m_{D^*}}}\right ]^2
\eeq

\beq
\label{equation81}
\left [ -4 \chi_2(1) + 12\chi '_3(1)\right  ]^2 \leq  {1 \over 3}  \left [ 5 \sigma^2 - 4 \rho^2 - 3(\rho^2)^2\right ] \sum_{n\not= 0} \left [ {1 \over \Delta E_{1/2}^{(n)}}{<D^{(n)} (v) |O_{mag}^{(c)} (0)|D(v)> \over \sqrt{4m_{D^{(n)}}m_D}}\right ]^2
\eeq

\noi Therefore, in the limit $\rho^2 \to {3 \over 4}$, $\sigma^2 \to {15 \over 16}$, one obtains
\beq
\label{equation82}
\chi ' _1(1) = \chi_2(1) =  \chi '_3(1) = 0
\eeq

\noi as it has been expected from the inspection of relations (\ref{67})-(\ref{69}).\par

This is a very strong correlation relating the behaviour of the elastic IW function $\xi (w)$ to the elastic subleading IW functions $\chi_i (w)$ ($i = 1,2,3$) near zero recoil.

\section{Discussion and phenomenological implications on the determination of $|{\bf V}_{\bf cb}|$.} \hspace*{\parindent} 
The bounds that relate second order subleading corrections
$\delta_{1/m^2}$, the first order $1/m_Q$ form factors $\chi_i(w)$ ($i
= 1,2,3$) and the curvature and slope of the elastic Isgur-Wise
function $\xi (w)$ should be taken into account in the exclusive determination of
$|V_{cb}|$.\par

On the one hand, the usual present point of view is that the exclusive
determination of $|V_{cb}|$ is not competitive with the inclusive
determination, that looks much more precise. However, one must keep in
mind that the hadronic uncertainties in both methods are of different
nature and that only a convergence of both can be satisfactory for a
precise measurement of $|V_{cb}|$.\par

As an illustration of the most advanced measurements, let us quote the
results of Babar \cite{17r}. To have  a qualitative feeling, let us add
the errors in quadrature,
\bea
\label{new77}
&&|V_{cb}|_{inclusive} = 0.0414 \pm 0.0008 \\
&&|V_{cb}|_{exclusive} = 0.0370 \pm 0.0020 
\label{new78}
\eea

\noi where the exclusive determination comes form $\overline{B} \to
D^*\ell \nu$ and uses the value 
\beq
\label{re85}
-\delta_{1/m^2}^{(A_1)} = 0.09 \pm 0.05
\eeq

\noi discussed in Appendix B.\par

The slight disagreement between both
determinations (\ref{new77}), (\ref{new78}) seems to suggest that $-\delta_{1/m^2}^{(A_1)}$ could be
larger than (\ref{re85}).\par

On the other hand, although this is not the main object of our
discussion, in obtaining $|V_{cb}|_{inclusive}$ one {\it fits} $\mu_G^2 = (0.27 \pm 0.07)$~GeV$^2$.
This is roughly within $1\sigma$ in agreement with the experimental
value obtained from the spectrum, namely $\mu_G^2 = 0.36$~GeV$^2$.
However, it seems to us that this parameter is a very well determined
quantity that, in the fit, should be fixed at this latter value. This
is just to emphasize that, even in the very efficient inclusive
determination, there are presumably still hadronic uncertainties.\par

Coming back to the exclusive determination, it is well known that there
is a great dispersion of the data in the different experiments using
$\overline{B} \to D(D^*)\ell\nu$, as discussed in detail by Grinstein and Ligeti \cite{newref}
(see also \cite{17r}).\par

Since in this determination, for example in $\overline{B} \to
D^*\ell\nu$, enters $-\delta_{1/m^2}^{(A_1)}$ and also the subleading
form factors $\chi_i (w)$ ($i = 1,2,3)$, as well as the shape of the
Isgur-Wise function $\xi (w)$, our bound (\ref{55e}) has to be taken into
account, as well as the vanishing of $\chi '_1(1)$, $\chi_2(1)$, $\chi '_3(1)$ in the limit $\rho^2 \to {3 \over 4}$, $\sigma^2 \to {15 \over 16}$.\par

The functions $\chi_i(w)$ $(i=1,2,3)$ have been computed in the
framework of the QCD Sum Rules approach \cite{12r} \cite{13r}
\cite{14r}, obtaining 
\bea
\label{56e}
&&\chi '_1(1) = (0.15 \pm 0.10)\ \overline{\Lambda} \nn \\
&&\chi_2(1) = - (0.05 \pm 0.01)\ \overline{\Lambda} \nn \\
&&\chi '_3 (1) = (0.009 \pm 0.004)\ \overline{\Lambda}\ .
\eea

\noi We have extracted these rough numbers from figures 5.5 of ref.
\cite{14r}, where the $\chi_i(w)$ ($i=1,2,3$) are dimensionless, given
in units of $\overline{\Lambda}$ and we have translated them in the definition  of ref. \cite{10r}, adopted in the present paper. On the other hand, one obtains, in the QCDSR approach 
\beq
\label{lasteq}
\rho_{ren}^2 \cong 0.7 
\eeq

Therefore, the QCDSR approach does not make explicit the constraint that we have obtained, and our discussion cannot proceed further within this scheme. \par

In the case of $\overline{B} \to D\ell \nu$ the correction
$-\delta_{1/m^2}^{(1)}$ is one of the pieces that constitute the
$1/m_Q^2$ correction~: besides $\ell_1$ there is another correction
$\ell_4$ \cite{18r} not concerned by our bounds, and therefore the
situation is less clear. Nevertheless, what we have said about $-
\delta_{1/m^2}^{(A_1)}$ applies to $- \delta _{1/m^2}^{(1)}$.\par

By considering his BPS limit, Uraltsev \cite{reference} has obtained complementary results. We will discuss separately the relation of his approach with our above sum rules.

\section{Conclusion.} \hspace*{\parindent}
To conclude, we have obtained bounds that relate $1/m_Q^2$ corrections
of form factors protected by Luke theorem, namely $h_{A_1}(w)$,
$h_+(w)$ and $h_1(w)$ to the $1/m_Q$ subleading form factors of the
Lagrangian type $\chi_i(w)$ ($i=1,2,3$) and to the shape of the elastic
Isgur-Wise $\xi (w)$. These bounds should in principle be taken into
account in the analysis of the exclusive determination of $|V_{cb}|$ in
the channels $\overline{B} \to D(D^*)\ell \nu$. On the other hand, we have demonstrated an important constraint on the behavior of the subleading form factors $\chi_i(w)$ in the limit
$\rho^2 \to {3 \over 4}$, $\sigma^2 \to {15 \over 16}$, since $\chi'_1(1)$, $\chi_2(1)$ and $\chi '_3(1)$ must vanish in this limit.\par

It would be very interesting to have a theoretical estimation of the functions $\chi_i(w)$ ($i = 1,2,3$) satisfying this constraint. Otherwise it seems questionable to try an exclusive determination of $|V_{cb}|$ by fitting the slope $\rho^2$ and considering uncorrelated subleading corrections, for example roughly constant values for $\chi '_1(1)$, $\chi_2(1)$ and $\chi'_3(1)$.
\vskip 1 truecm 

\section*{Appendix A. Derivation of the Lagrangian Sum Rules using the OPE.}
\hspace*{\parindent} In this Appendix we give an alternative derivation
of the SR (\ref{26e})-(\ref{28e}), following the same method used in
ref. \cite{7r}, based on the OPE, to obtain similar SR concerning the
$1/m_Q$ perturbations of the heavy quark current. \par

To make easier the study of the subleading corrections we did
consider the following limit
$$m_c \gg m_b \gg \Lambda_{QCD} \ . \eqno({\rm A.1})$$

\noi Then, as explained in \cite{7r}, the difference between the two
energy denominators in the $T$-product (\ref{1e}) is large
$$q^0 - E_f + E_{X_{\overline{c}bb}}- \left ( q^0 + E_i - E_{X_c} \right )  \sim 2m_c \eqno({\rm A.2})$$

\noi where $X_c$ and $X_{\overline{c}bb}$ denote the intermediate
states of the direct and $Z$ orderings. Therefore, we can in this limit
neglect the $Z$ diagram, and consider the imaginary part of the direct
diagram, the piece proportional to 
$$\delta \left ( q^0 + E_i - E_{X_c}\right ) \ .\eqno({\rm A.3})$$

\noi Notice that one can choose $q^0$ such that there is a left-hand
cut, even in the conditions (A.1). This means that $q^0$ is of the
order of $m_c$ and $m_c - q^0$ is fixed, of the order $m_b$. Our
conditions are, in short, as follows~: 
$$\Lambda_{QCD} \ll m_b \sim m_c - q^0 \ll q^0 \sim m_c \to \infty\ . \eqno({\rm A.4})$$ 

\noi To summarize, we did consider the heavy quark limit for the $c$
quark, but allowing for a large finite mass for the $b$ quark.\par

The final result is the sum rule \cite{7r}
$$\sum_{D_n} <B_f (v_f) |J_f (0)|D_n (v')>\ <D_n(v')|J_i(0)|B_i(v_i)>$$
$$=\ <B(v_f)|\overline{b}(0)\Gamma_f {1 + {/ \hskip - 2 truemm v}' \over 2v'^0} \Gamma_i b(0) |B (v_i) > + \ O(1/m_c) \eqno({\rm A.5})$$

\noi that is valid for {\it all powers} of an expansion in $1/m_b$, but
only to leading order in $1/m_c$.\par

At leading order $m_b , m_c \to \infty$ one gets the SR formulated in \cite{3r}-\cite{5r}. In
ref. \cite{7r} we considered the first order in $1/m_b$ to both the
left and right hand sides of (A.5), using the formalisms of Falk and
Neubert \cite{18r} for the ${1 \over 2}^- \to {1 \over 2}^-$
transitions and of Leibovich et al. \cite{8r} for the ${1 \over 2}^-
\to {1 \over 2}^+$, ${1 \over 2}^- \to {3 \over 2}^+$ transitions. The
formalism was extended to all possible transitions ${1 \over 2}^- \to
j^{\pm}$ \cite{9r}.\par

We did consider only the $1/m_Q$ perturbations that are perturbations
of the current, namely $L_4(w)$, $L_5(w)$ and $L_6(w)$, in the notation
of \cite{18r}. To obtain the maximum information we
did consider in \cite{7r} initial and final pseudoscalar $B(v_i)
\to B(v_f)$ or vector states $B^*(v_i, \varepsilon_i) \to B^*(v_f, \varepsilon_f)$. This yielded to
two interesting very simple sum rules. The reason is that
we considered the SR at the frontier
$$\left ( w_i, w_f, w_{if}\right ) = (w, 1, w) \eqno({\rm A.6})$$

\noi of the domain of the variables $(w_i, w_f, w_{if}) = (v_i \cdot
v', v_f \cdot v', v_i\cdot v_f)$ \cite{3r},
$$ w_i \geq 1 \qquad \qquad w_f \geq 1$$
$$w_iw_f - \sqrt{(w_i^2 - 1) (w_f^2-1)}\ \leq w_{if} \leq w_i w_f + \sqrt{(w_i^2 - 1) (w_f^2 - 1)} \ . \eqno({\rm A.7})$$

In this Appendix we formulate new SR for the {\it Lagrangian}
perturbations, parallel to the ones on the {\it Current} perturbations
(\ref{3e})-(\ref{4e}), using the OPE formalism of ref. \cite{7r}. We
find the same results than with the simple method exposed in detail in
Section 2.\par

In obtaining (\ref{3e})-(\ref{4e}) we did use axial currents aligned
along the initial and final velocities, $\Gamma_i = {/\hskip - 2 truemm
v}_i \gamma_5$, $\Gamma_f = {/\hskip - 2 truemm v}_f \gamma_5$. Let us
use now the {\it vector} heavy quark currents, aligned along the
initial and final four-velocities,
$$\Gamma_i = {/\hskip - 2 truemm v}_i \qquad \qquad \qquad \Gamma_f = {/\hskip - 2 truemm v}_f \eqno({\rm A.8})$$

\noi and proceed as in \cite{7r}. Using expression (A.5) we obtain two
sum rules at order $1/m_b$ for initial and final pseudoscalar $B(v_i)
\to B(v_f)$ or vector states $B^*(v_i, \varepsilon_i) \to B^*(v_f, \varepsilon_f)$. \par

To compute the SR we need the following matrix elements, including the
$1/m_b$ order \cite{18r}
$$<D(v')|\overline{Q}'\Gamma Q|B(v)>\ = - \xi (w) Tr [\overline{D}(v') \Gamma B(v)]$$
$$- {1 \over 2m_b} Tr \left \{ \overline{D} (v') \Gamma \left [ P_+L_+(v, v') + P_- L_- (v, v') \right ] \right \} + O(1/m_c)\ . \eqno({\rm A.9})$$

The $4 \times 4$ matrices write, for a pseudoscalar meson $M$ (initial $B$ or intermediate $D$)
$$M(v) = P_+ (v) (-\gamma_5)$$
$$P_+(v)L_+(v,v') + P_- (v) L_- (v, v') = \left [ L_1(w) P_+(v) + L_4(w) P_-(v) \right ] (-\gamma_5) \eqno({\rm A.10})$$

\noi where $w = v \cdot v'$, while for a vector meson $M$ one has 
$$M(v) = P_+ (v) {/\hskip - 2 truemm \varepsilon }$$
$$P_+(v) L_+(v, v') + P_- (v) L_-(v,v') =$$
$$P_+(v) \left [ {/\hskip - 2 truemm \varepsilon } L_2(w) + (\varepsilon_v \cdot v')L_3(w) \right ] + P_- (v) \left [ {/\hskip - 2 truemm \varepsilon } L_5(w) + (\varepsilon \cdot v')L_6(w) \right ] \ . \eqno({\rm A.11})$$

The matrix elements to excited states write \cite{8r}
$$<D\left ( \textstyle{{3 \over 2}^+}\right ) (v') |\overline{c}\Gamma b|B(v)>\ = \sqrt{3}\ \tau_{3/2} (w) \ Tr \left [ v_{\sigma} \overline{D}^{\sigma} (v') \Gamma B(v) \right ]$$
$$+ {1 \over 2m_b} \left \{ Tr \left [ S_{\sigma \lambda}^{(b)} \overline{D}^{\sigma} (v') \Gamma \gamma^{\lambda} B(v) \right ] + \eta_{ke}^{(b)} Tr \left [ v_{\sigma} \overline{D}^{\sigma} (v') \Gamma B(v) \right ]\right .$$
$$\left . + \ Tr \left [ R_{\sigma\alpha\beta}^{(b)} \overline{D}^{\sigma}(v') \Gamma P_+ (v) i \sigma^{\alpha\beta} B(v) \right ] \right \} + O(1/m_c) $$

$$<D\left ( \textstyle{{1 \over 2}^+}\right ) (v') |\overline{c}\Gamma b|B(v)>\ = 2\tau_{1/2} (w) \ Tr \left [ \overline{D}(v') \Gamma B(v) \right ]$$
$$+ {1 \over 2m_b} \left \{ Tr \left [ S_{\lambda}^{(b)} \overline{D}(v') \Gamma \gamma^{\lambda} B(v) \right ] + \chi_{ke}^{(b)} Tr \left [ \overline{D} (v') \Gamma B(v) \right ]\right .$$
$$\left . + \ Tr \left [ R_{\alpha\beta}^{(b)} \overline{D}(v') \Gamma P_+ (v) i \sigma^{\alpha\beta} B(v) \right ] \right \} + O(1/m_c) \eqno({\rm A.12})$$

\noi where
$$D_{2^+}^{\sigma}(v') = P_+ (v') \varepsilon_{v'}^{\sigma\nu} \gamma_{\nu}$$
$$D_{1^+}^{\sigma}(v') = - \sqrt{{3 \over 2}} P_+ (v') \varepsilon_{v'}^{\nu} \gamma_{5}\left [ g_{\nu}^{\sigma} - {1 \over 3} \gamma_{\nu} (\gamma^{\sigma} - v'^{\sigma})\right ]$$
$$D_{1^+}(v') = P_+ (v') \varepsilon_{v'}^{\nu} \gamma_5\gamma_{\nu}$$
$$D_{0^+}(v') = P_+ (v') \ . \eqno({\rm A.13})$$

\noi The notation $S_{\sigma\lambda}^{(b)}$, $S_{\lambda}^{(b)}$ denote
the perturbations to the current, and $\eta_{ke}^{(b)}$,
$\chi_{ke}^{(b)}$ and $R_{\sigma \alpha\beta}^{(b)}$,
$R_{\alpha\beta}^{(b)}$ denote respectively the kinetic and the magnetic Lagrangian
perturbations.\par

Expanded in terms of Lorentz covariant factors and subleading IW
functions, these tensor quantities read \cite{8r}
$$\hskip - 1 truecm S_{\sigma\lambda}^{(Q)} = v_{\sigma} \left [ \tau_1^{(Q)}(w) v_{\lambda} + \tau_2^{(Q)}(w) v'_{\lambda} + \tau_3^{(Q)}(w) \gamma_{\lambda}\right ] + \tau_4^{(Q)}(w) g_{\sigma\lambda}$$
$$S_{\lambda}^{(Q)} = \zeta_1^{(Q)}(w) v_{\lambda} + \zeta_2^{(Q)}(w) v'_{\lambda} + \zeta_3^{(Q)}(w) \gamma_{\lambda} \qquad\qquad (Q = b, c) \eqno({\rm A.14})$$

\noi for the Current perturbations, and
$$R_{\sigma\alpha\beta}^{(b)} = \eta_1^{(b)} (w) v_{\sigma} \gamma_{\alpha} \gamma_{\beta} + \eta_2^{(b)}(w) v_{\sigma} v'_{\alpha} \gamma_{\beta} + \eta_3^{(b)} (w) g_{\sigma\alpha} v'_{\beta}$$
$$\hskip - 4 truecm R_{\alpha\beta}^{(b)} = \chi_1^{(b)} (w) \gamma_{\alpha} \gamma_{\beta} + \chi_2^{(b)}(w) v'_{\alpha} \gamma_{\beta} \eqno({\rm A.15})$$

\noi for the Lagrangian magnetic perturbations.\par

We have also to consider the intermediate states $D\left ( {3 \over
2}^-, 1^-\right )$, $D\left ( {3 \over 2}^-, 2^-\right )$. The
corresponding $4 \times 4$ matrices for the ${3 \over 2}^-$ states will
be given in terms of those of ${3 \over 2}^+$ states (A.13) by \cite{9r}
$$D_{1^-}^{\sigma}(v') = D_{1^+}^{\sigma}(v') (-\gamma_5) \ , \qquad D_{2^-}^{\sigma}(v') = D_{2^+}^{\sigma}(v') (-\gamma_5) \eqno({\rm A.16})$$ 

\noi and the current matrix elements, including $1/m_b$ corrections are
$$<D\left ( \scriptstyle{{3 \over 2}^-}\right ) (v') |\overline{c}\Gamma b|B(v)>\ = \tau_{3/2}^{(2)} (w) \ Tr \left [ v_{\sigma}\overline{D}^{\sigma}(v') \Gamma B(v) \right ]$$
$$+ {1 \over 2m_b} \left \{ Tr \left [ T_{\sigma\lambda}^{(b)} \overline{D}^{\sigma}(v') \Gamma \gamma^{\lambda} B(v) \right ] + \rho_{ke}^{(b)} \ Tr \left [ v_{\sigma}\overline{D}^{\sigma} (v') \Gamma B(v) \right ]\right .$$
$$\left . + \ Tr \left [ V_{\sigma\alpha\beta}^{(b)} \overline{D}^{\sigma}(v') \Gamma P_+ (v) i \sigma^{\alpha\beta} B(v) \right ] \right \} + O(1/m_c) \eqno({\rm A.17})$$

\noi where
$$T_{\sigma\lambda}^{(b)} = v_{\sigma} \left [ \sigma_1^{(b)}(w) v_{\lambda} + \sigma_2^{(b)} (w) v'_{\lambda} + \sigma_3^{(b)}(w) \gamma_{\lambda}\right ] + \sigma_4^{(b)}(w) g_{\sigma\lambda} \eqno({\rm A.18})$$

\noi denotes the Current perturbations, and
$$V_{\sigma\alpha\beta}^{(b)} = \rho_1^{(b)}(w) v_{\sigma} \gamma_{\alpha} \gamma_{\beta} + \rho_2^{(b)}(w) v_{\sigma} v'_{\alpha} \gamma_{\beta} + \rho_3^{(b)}(w) g_{\sigma\alpha} \gamma_{\beta} \eqno({\rm A.19})$$

\noi the corresponding Lagrangian perturbations. In defining (A.19) we
perform a different Lorentz decomposition as done in \cite{8r} for the
${1 \over 2}^+$, ${3 \over 2}^+$ states. The necessity of this
alternative parametrization is explained in ref. \cite{19r}. \par

Proceeding like in ref. \cite{7r}, starting from the master formula
(A.5), i.e. taking the formal limit $m_c \gg m_b$, and using now the
vector currents (A.8), we find the following sum rules for 
$$\left ( w_i, w_f, w_{if}\right ) = (w, 1, w) \qquad \hbox{or} \qquad \left ( v_i, v_f, v\right ) = (v, v', v') \ , \eqno({\rm A.20})$$

\noi respectively for the pseudoscalar $B(v) \to B(v')$ or $B^*(v, \varepsilon) \to
B^*(v', \varepsilon ')$ transitions, 
$$L_1(w) = \sum_n \xi^{(n)}(w) L_1^{(n)}(1) \eqno({\rm A.21})$$ 
$$L_2(w) + (w-1)L_3(w) = \sum_n \xi^{(n)}(w) L_2^{(n)}(1) - {2 \over 3} (w-1) \sum_n \tau_{3/2}^{(2)(n)}(w) \rho_3^{(n)}(1) \eqno({\rm A.22})$$  

In (A.21) and (A.22) one has a relation between the elastic subleading
form factors of Lagrangian type $L_1(w)$, $L_2(w)$ and $L_3(w)$ and
excited leading IW functions $\xi^{(n)}(w)$, $\tau_{3/2}^{(2)(n)}(w)$
and excited subleading form factors of Lagrangian type at zero recoil,
$L_1^{(n)}(1)$, $L_2^{(n)}(1)$ and $\rho_3^{(n)}(1)$. Notice that in the sums (A.21) and
(A.22) the terms $\xi^{(0)}(w)L_1^{(0)}(1)$, $\xi^{(0)}(w)L_1^{(0)}(1)$
do not contribute due to Luke theorem \cite{10r}
$$L_1(1) = L_1(1) = 0 \ . \eqno({\rm A.23})$$

\noi Therefore the SR (A.21), (A.22) actually reduce to
$$L_1(w) = \sum_{n\not= 0} \xi^{(n)}(w) L_1^{(n)}(1) \eqno({\rm A.24})$$ 
$$L_2(w) + (w-1)L_3(w) = \sum_{n\not= 0} \xi^{(n)}(w) L_2^{(n)}(1) - {2 \over 3} (w-1) \sum_n \tau_{3/2}^{(2)(n)}(w) \rho_3^{(n)}(1) \eqno({\rm A.25})$$  

We see that $L_1(w)$ and $L_2(w)$ satisfy Luke theorem at $w=1$, $L_1(1) = L_2(1) = 0$, due to
$\xi^{(n)}(1) = \delta_{n,0}$.\par

A number of comments are worth here to be added.\par

(i) All current perturbation form factors, the elastic  $L_4(w)$, $L_5(w)$ and
$L_6(w)$ and the inelastic ones cancel in the sum rules. Only perturbations of the
Lagrangian remain.\par

(ii) Only the ${1 \over 2}^-$ and ${3 \over 2}^-$ intermediate states
contribute at the frontier (A.20). \par

(iii) The SR (A.24)-(A.25) are reminiscent of the SR
(\ref{3e})-(\ref{4e}), that relate elastic subleading form factors of
the current type to leading order excited IW functions and subleading
excited form factors at zero recoil. In this case, however, these latter
form factors can be simply expressed, by the equations of motion, in terms of leading IW functions and
level spacings.\par

(iv) It can be easily shown, following the same type of arguments as in
\cite{7r} that higher excited states do not contribute to the SR
(A.24)-(A.25) because we choose the frontier (A.20).

(v) Notice that for the SR concerning $1/m_Q$ perturbations to the {\it
Current}, only ${1 \over 2}^+$ and  ${3 \over 2}^+$ intermediate states
survive. Similarly, in a symmetric way, for the $1/m_Q$ perturbations of the {\it
Lagrangian}, only ${1 \over 2}^-$ and ${3 \over 2}^-$ intermediate
states survive. \par

Writing the combinations $L_1(w)$ and $L_2(w) + (w-1)L_3(w)$ in terms
of the ${\cal L}_{kin}$ and ${\cal L}_{mag}$ matrix elements $\chi_i(w)$
\cite{18r} \cite{10r},
$$L_1(w) = 2\chi_1(w) - 4(w-1) \chi_2(w) + 12 \chi_3(w)$$
$$\hskip - 3.5 truecm L_2(w) = 2\chi_1(w) - 4 \chi_3(w)$$
$$\hskip - 5 truecm L_3(w) = 4 \chi_2(w)  \eqno({\rm A.26})$$

\noi We realize that, as obtained in Section 2, the combination
$L_1(w)$ gets contributions only from ${1\over 2}^-$ intermediate states,
while the combination $L_2(w)+(w-1)L_3(w) = 2\chi_1(w) + 4(w-1) \chi_2
(w) - 4\chi_3(w)$ contains contributions from ${1 \over 2}^-$ and ${3
\over 2}^-$ intermediate states, as we have found in Section 3. In terms
of the $\chi_i(w)$ ($i = 1,2,3)$, the SR write

$$\hskip - 5 truecm 2\chi_1(w) - 4(w-1) \chi_2 (w) + 12 \chi_3(w) = \sum_{n\not= 0} \xi^{(n)}(w) L_1^{(n)}(1)$$
$$2\chi_1(w) + 4(w-1) \chi_2 (w) -4 \chi_3(w) = \sum_{n\not= 0} \xi^{(n)}(w) L_2^{(n)}(1) - {2 \over 3} (w-1) \sum_n \tau_{3/2}^{(2)(n)} (w) \rho_3^{(n)}(1)  \eqno({\rm A.27})$$

\noi On the other hand, since only ${\cal L}_{kin}$ contributes to
$\chi_1(w)$ and to $\chi_1^{(n)}(1)$, decomposing $L_1^{(n)}(1)$ and
$L_2^{(n)}(1)$ in terms of $\chi_i^{(n)}(1)$ like in the first two
relations (A.26), we can solve for $\chi_2(w)$ and $\chi_3(w)$ and find
finally

$$\hskip - 7 truecm \chi_1(w) = \sum_{n \not= 0} \xi^{(n)}(w) \chi_1^{(n)}(1)$$
$$\hskip - 6 truecm \chi_2(w) = - {1 \over 4} \sum_n \tau_{3/2}^{(2)(n)}(w) \rho_3^{(n)}(1)$$
$$\chi_3(w) = \sum_{n \not= 0} \xi^{(n)}(w) \chi_3^{(n)}(1) - {1 \over 12}(w-1) \sum_n \tau_{3/2}^{(2)(n)}(w) \rho_3^{(n)}(1)\eqno({\rm A.28})$$

 From the definition of $\chi_1^{(n)}(1)$, $\chi_3^{(n)}(1)$ and
$\rho_3^{(n)}(1)$ from the $T$-products as in (\ref{5e}) and (\ref{6e})
for ${1 \over 2}^- \to {1 \over 2}^-$ transitions, but allowing for $n
\not= 0$, and the corresponding one for ${1 \over 2}^- \to {3 \over
2}^-$ transitions, we see that these relations are identical with
(\ref{26e})-(\ref{28e}), obtained in Section 2 just from
the definition of the form factors $\chi_i(w)$. The
inelastic form factors at zero recoil $\chi_1^{(n)}(1)$ ($n \not= 0$) are
given by the matrix elements ${1 \over 2}^-$ ($n = 0$) $\to {1 \over
2}^-$ ($n \not= 0$) of ${\cal L}_{kin}$ ponderated by the corresponding
energy denominators. Similarly, $\chi_3^{(n)}(1)$ and $\rho_3^{(n)}(1)$
($n \geq 0$) are given by the matrix elements ${1\over 2}^-$ ($n = 0$)
$\to {1 \over 2}^-$ ($n \not= 0$) and ${1\over 2}^-$ ($n = 0$)
$\to {3 \over 2}^-$ ($n \geq 0$) coming from ${\cal L}_{mag}$.

\section*{Appendix B.}
\hspace*{\parindent}
Although the QCDSR results (\ref{56e}) do not explicitely satisfy the constraints (\ref{equation82}), it could be of some interest to use these results to estimate the r.h.s. of  (\ref{67})-(\ref{69}) varying the input for the slope $\rho^2$ and the curvature $\sigma^2$. The aim would be to see how these bounds evolve as one approaches the limit $\rho^2 \to {3 \over 4}$, $\sigma^2 \to {15 \over 16}$.\par

We denote the bounds under the form of the contribution of
the OPE term of the matrix elements (\ref{34e}), plus the ${1\over 2}^-$ and ${3 \over 2}^-$ inelastic
contributions,
$$- \delta_{1/m^2}^{(A_1)} \geq \hbox{OPE} + {1 \over 2}^- + {3 \over 2}^-  \ . \eqno({\rm B.1})$$

\noi In view of the theoretical comments on the bounds made in the
preceding Section, we can only provide some qualitative numerical
illustrations that will show the general trend of the results. We give
some numerical results in Tables 1, 2 and 3 using the parameters
$$m_c = 1.25 \ \hbox{GeV} \qquad \qquad m_b = 4.75\ \hbox{GeV} \qquad \qquad \overline{\Lambda} = 0.50\ \hbox{GeV}
$$
$$\mu_{\pi}^2 = 0.50\ \hbox{GeV}^2 \qquad \qquad \mu_G^2 = 0.36\ \hbox{GeV}^2 
\eqno({\rm B.2})$$

\noi and for the curvature $\sigma^2$ of the Isgur-Wise function we use its
value in terms of the slope $\rho^2$ given by the ``dipole'' Ansatz
\cite{15r}
$$\xi (w) = \left ( {2 \over w+1}\right )^{2 \rho^2} \eqno({\rm B.3})$$

\noi namely
$$\sigma^2 = {\rho^2 \over 2} + (\rho^2)^2\ .
\eqno({\rm B.4})$$

\noi We use this relation between the curvature and the slope because, as shown in \cite{5r},
(B.3) satisfies all the bounds that we have obtained for the
derivatives of the elastic IW function \cite{3r} \cite{4r} \cite{5r}.

\vskip 5 truemm
\begin{center}
\begin{tabular}{|l|l|}
\hline
Parameters &$- \delta_{1/m^2}^{(A_1)} \geq \hbox{OPE} + {1 \over 2}^- + {3 \over 2}^-$ \\
 \hline
 (i) $\rho^2 = 1$ \qquad \quad $\sigma^2 = 1.5$ & \\
 $\chi '_1(1) = 0.15\ \overline{\Lambda}$ &$-\delta_{1/m^2}^{(A_1)}$\\
 $\chi_2 (1) = - 0.05\ \overline{\Lambda}$ &$\geq 0.052 + 0.000 + 0.003$ \\
 $\chi '_3(1) = 0.01\ \overline{\Lambda}$ &$= 0.055$\\
\hline
 (ii)  $\rho^2 = 1$ \qquad \quad $\sigma^2 = 1.5$ & \\
 $\chi '_1(1) = 0.15\ \overline{\Lambda}$ &$-\delta_{1/m^2}^{(A_1)}$\\
 $\chi_2 (1) = - 0.05\ \overline{\Lambda}$ &$\geq 0.052 + 0.113 + 0.003$ \\
 $\chi '_3(1) = 0.15\ \overline{\Lambda}$ &$= 0.168$\\
 \hline
(iii)  $\rho^2 = 0.9$ \qquad $\sigma^2 = 1.26$ & \\
$\chi '_1(1) = 0.15\ \overline{\Lambda}$ &$-\delta_{1/m^2}^{(A_1)}$\\
 $\chi_2 (1) = - 0.05\ \overline{\Lambda}$ &$\geq 0.052 + 0.000 + 0.005$ \\
 $\chi '_3(1) = 0.01\ \overline{\Lambda}$ &$= 0.057$\\
\hline
 (iv)  $\rho^2 = 0.8$ \qquad $\sigma^2 = 1.04$ & \\
  $\chi '_1(1) = 0.15\ \overline{\Lambda}$ &$-\delta_{1/m^2}^{(A_1)}$\\
 $\chi_2 (1) = - 0.05\ \overline{\Lambda}$ &$\geq 0.001 + 0.001 + 0.017$ \\
 $\chi '_3(1) = 0.01\ \overline{\Lambda}$ &$= 0.070$\\
 \hline
 (v) $\rho^2 = 0.76$ \qquad $\sigma^2 = 0.96$ & \\
 $\chi '_1(1) = 0.15\ \overline{\Lambda}$ &$-\delta_{1/m^2}^{(A_1)}$\\
 $\chi_2 (1) = - 0.05\ \overline{\Lambda}$ &$\geq 0.052 + 0.004 + 0.088$ \\
 $\chi '_3(1) = 0.01\ \overline{\Lambda}$ &$= 0.14$\\
 \hline
\end{tabular}
\end{center}

\noi {\bf Table 1.} The lower bound (\ref{55e}) for
$-\delta_{1/m^2}^{(A_1)}$ for different values of the parameters. OPE
denotes the contribution depending on $\mu_{\pi}^2$, $\mu_G^2$ and
${1\over 2}^-$, ${3 \over 2}^-$ the inelastic contributions of the
corresponding $1^-$ excited states. We fix the values $m_c = 1.25$~GeV,
$m_b = 4.75$~GeV, $\overline{\Lambda} = 0.5$~GeV, $\mu_{\pi}^2 =
0.50$~GeV$^2$, $\mu_G^2 = 0.36$~GeV$^2$.\par \vskip 5 truemm

An order of magnitude estimation of the r.h.s. of relation (\ref{36e})
assumes that the inelastic term is roughly a factor $\chi$ of the OPE result \cite{16r}, i.e.
$$- \delta_{1/m^2}^{(A_1)} = (1 + \chi ) \left [ {\mu_G^2 \over 6m_c^2} + {\mu_{\pi}^2 - \mu_{G}^2 \over 8} \left ( {1 \over m_c^2} + {1 \over m_b^2} + {2 \over 3m_cm_b}\right ) \right ] \ .
\eqno({\rm B.5})$$
\vskip 3 truemm

\noi Taking $\chi = 0.5 \pm 0.5$ \cite{16r}, one gets, for $\mu_{\pi}^2 = 0.50$, $\mu_G^2 = 0.36$, 
$$- \delta_{1/m^2}^{(A_1)} \cong 0.09 \pm 0.05\ .
\eqno({\rm B.6})$$

We observe from the results of Table 1 that the lower bound grows
rapidly as one approaches the lower bounds $\rho^2 = {3 \over 4}$, $\sigma^2 = {15 \over 16}$.
However, for the values chosen for $\rho^2$, $\sigma^2$, the guess (B.6) can
be accomodated with the QCDSR estimations for $\chi_i(w)$ $(i =
1,2,3)$.\par

Let us comment on the different entries of Table 1. Our results can
only pretend to give the qualitative trend of the bounds. In the choice
of parameters (i) we have used the central values (\ref{56e}) and $\rho^2 =1$. The lower bound on $- \delta_{1/m^2}^{(A_1)}$ is
dominated by the OPE contribution, and specially the ${1\over 2}^-$
contributions are very small because of a strong cancellation between two terms in (\ref{55e}). In the second row (ii), just as an illustration, we have taken
the central values of (\ref{56e}) except for $\chi '_3(1)$, for which
we have taken the large value suggested by Grinstein and Ligeti
\cite{newref} to fit the different experiments on $\overline{B} \to
D(D^*)\ell \nu$, keeping however $\rho^2 = 1$. We observe that now the ${1\over
2}^-$ contribution becomes very large. In choices (iii), (iv) and
(v) we take still the central values of (\ref{56e}) and we
decrease the value of $\rho^2 = 0.9$, $0.8$, $0.76$, and consequently
the curvature. For (v) the inelastic contributions
become sizeable, specially for the ${3 \over 2}^-$ contributions. Of
course, the bounds diverge for $\rho^2 = {3 \over 4}$, $\sigma^2 = {15 \over 16}$. This value for $\rho^2$ is
not far away from the QCDSR value $\rho^2_{ren} = 0.7$. However,  strictly speaking, we cannot make a comparison because we do not have computed the
radiative corrections to our bound. The same comment applies to the
functions $\chi_i(w)$ computed in the QCDSR approach. Therefore, our
numerical results can only be indicative of what can be expected.\par

In Tables 2 and 3 we give the lower bounds on $- \ell_1(1)$ and $-\ell_2(1)$ and the corresponding lower
bounds on $- \delta_{1/m^2}^{(+)}$, $- \delta_{1/m^2}^{(1)}$, using the
same sets of parameters as in Table 1.\par

\vskip 5 truemm
\begin{center}
\begin{tabular}{|l|l|l|}
\hline
Parameters &$- \ell_1 (1) \geq \hbox{OPE} +  {1 \over 2}^- + {3 \over 2}^-$ &$- \delta_{1/m^2}^{(+)} \geq \hbox{OPE} +  {1 \over 2}^- + {3 \over 2}^-$ \\
 \hline
 (i) $\rho^2 = 1$ \qquad \quad $\sigma^2 = 1.5$ & &\\
$\chi '_1 (1) = 0.15\ \overline{\Lambda}$ &$-\ell_1(1) \geq$ &$-\delta_{1/m^2}^{(+)}$\\
$\chi_2 (1) = - 0.05\ \overline{\Lambda}$ &$(0.075 + 0.288 + 0)$ GeV$^2$   &$\geq 0.007 + 0.025 + 0$ \\ 
 $\chi '_3(1) = 0.01\ \overline{\Lambda}$  &$=0.363$ GeV$^2$ &$=0.032$ \\
 \hline
  (ii) $\rho^2 = 1$ \qquad \quad $\sigma^2 = 1.5$ & &\\
$\chi '_1 (1) = 0.15\ \overline{\Lambda}$ &$-\ell_1(1) \geq$ &$-\delta_{1/m^2}^{(+)}$\\
 $\chi_2 (1) = - 0.05\ \overline{\Lambda}$ &$(0.075 + 3.967 + 0)$ GeV$^2$ &$\geq 0.007 + 0.345 + 0$ \\  
 $\chi '_3(1) = 0.15\ \overline{\Lambda}$ &$=4.042$ GeV$^2$ &$=0.351$ \\
 \hline
(iii) $\rho^2 = 0.9$ \qquad $\sigma^2 = 1.26$ & &\\
 $\chi '_1 (1) = 0.15\ \overline{\Lambda}$ &$-\ell_1(1) \geq$ &$-\delta_{1/m^2}^{(+)}$\\
 $\chi_2 (1) = - 0.05\ \overline{\Lambda}$ &$(0.075 + 0.534 + 0)$ GeV$^2$ &$\geq 0.007 + 0.046 + 0$ \\ 
 $\chi '_3(1) = 0.01\ \overline{\Lambda}$ &$=0.609$ GeV$^2$ &$=0.053$ \\
 \hline
 (iv) $\rho^2 = 0.8$ \qquad $\sigma^2 = 1.04$ & &\\
 $\chi '_1 (1) = 0.15\ \overline{\Lambda}$ &$-\ell_1(1) \geq$ &$-\delta_{1/m^2}^{(+)}$\\
 $\chi_2 (1) = - 0.05\ \overline{\Lambda}$ &$(0.075 + 1.802 + 0)$ GeV$^2$ &$\geq 0.007 + 0.156 + 0$ \\ 
 $\chi '_3(1) = 0.01\ \overline{\Lambda}$ &$=1.877$ GeV$^2$ &$=0.163$ \\
 \hline
 (v)  $\rho^2 = 0.76$ \qquad $\sigma^2 = 0.96$ & &\\
 $\chi '_1 (1) = 0.15\ \overline{\Lambda}$ &$-\ell_1(1) \geq$ &$-\delta_{1/m^2}^{(+)}$\\
 $\chi_2 (1) = - 0.05\ \overline{\Lambda}$ &$(0.075 + 9.484 + 0)$ GeV$^2$ &$\geq 0.007 + 0.824 + 0$ \\  
 $\chi '_3(1) = 0.01\ \overline{\Lambda}$ &$=9.559$ GeV$^2$ &$=0.831$ \\
\hline
 \end{tabular}
\end{center}
\noi {\bf Table 2.} The lower bounds for $-\ell_1(1)$ (\ref{69e}) and $-\delta_{1/m^2}^{(+)}$ (\ref{68}) for the same set of parameters and notations of Table 1. \par
\vskip 5 truemm

\begin{center}
\begin{tabular}{|l|l|l|}
\hline
Parameters &$- \ell_2 (1) \geq \hbox{OPE} +  {1 \over 2}^- + {3 \over 2}^-$ &$- \delta_{1/m^2}^{(1)} \geq \hbox{OPE} +  {1 \over 2}^- + {3 \over 2}^-$ \\
 \hline
 (i) $\rho^2 = 1$ \qquad \quad $\sigma^2 = 1.5$ & &\\
   $\chi '_1 (1) = 0.15\ \overline{\Lambda}$ &$-\ell_2(1) \geq$ &$-\delta_{1/m^2}^{(1)}$\\
 $\chi_2 (1) = - 0.05\ \overline{\Lambda}$  &$(0.308 + 0.028 + 0.017)$ GeV$^2$ &$\geq 0.027 + 0.002 + 0.001$ \\ 
 $\chi '_3(1) = 0.01\ \overline{\Lambda}$ &$=0.353$ GeV$^2$ &$=0.030$ \\
 \hline
  (ii)  $\rho^2 = 1$ \qquad\quad  $\sigma^2 = 1.5$ & &\\  
$\chi '_1 (1) = 0.15\ \overline{\Lambda}$ &$-\ell_2(1) \geq$ &$-\delta_{1/m^2}^{(1)}$\\
 $\chi_2 (1) = - 0.05\ \overline{\Lambda}$ &$(0.308 + 0.101 + 0.017)$ GeV$^2$ &$\geq 0.027 + 0.009 + 0.001$ \\ 
 $\chi '_3(1) = 0.15\ \overline{\Lambda}$ &$=0.426$ GeV$^2$ &$=0.037$ \\
\hline
 (iii)  $\rho^2 = 0.9$ \qquad $\sigma^2 = 1.26$ & &\\
$\chi '_1 (1) = 0.15\ \overline{\Lambda}$ &$-\ell_2(1) \geq$ &$-\delta_{1/m^2}^{(1)}$\\
 $\chi_2 (1) = - 0.05\ \overline{\Lambda}$ &$(0.308 + 0.052 + 0.031)$ GeV$^2$ &$\geq 0.027 + 0.004 + 0.003$ \\ 
 $\chi '_3(1) = 0.01\ \overline{\Lambda}$ &$=0.391$ GeV$^2$ &$=0.034$ \\
 \hline
 (iv) $\rho^2 = 0.8$ \qquad $\sigma^2 = 1.04$ & &\\
 $\chi '_1 (1) = 0.15\ \overline{\Lambda}$ &$-\ell_2(1) \geq$ &$-\delta_{1/m^2}^{(1)}$\\
 $\chi_2 (1) = - 0.05\ \overline{\Lambda}$ &$(0.308 + 0.175 + 0.104)$ GeV$^2$ &$\geq 0.027 + 0.015 + 0.009$ \\ 
 $\chi '_3(1) = 0.01\ \overline{\Lambda}$ &$=0.587$ GeV$^2$ &$=0.051$ \\
 \hline
 (v)  $\rho^2 = 0.76$ \qquad $\sigma^2 = 0.96$ & &\\
$\chi '_1 (1) = 0.15\ \overline{\Lambda}$ &$-\ell_2(1) \geq$ &$-\delta_{1/m^2}^{(1)}$\\
 $\chi_2 (1) = - 0.05\ \overline{\Lambda}$ &$(0.308 + 0.922 + 0.548)$ GeV$^2$ &$\geq 0.027 + 0.080 + 0.048$ \\ 
 $\chi '_3(1) = 0.01\ \overline{\Lambda}$ &$=1.778$ GeV$^2$ &$=0.155$ \\
 \hline
 \end{tabular}
\end{center}
\noi {\bf Table 3.} The lower bounds for $-\ell_2(1)$ (\ref{70e}) and for $-\delta_{1/m^2}^{(1)}$ (\ref{69}) for the same set of parameters and notations of Table 1. \par
\vskip 5 truemm

Concerning Table 2, we observe that
the bounds on $-\ell_1(1)$ and on $-\delta_{1/m^2}^{(+)}$ are now dominated
by the ${1\over 2}^-$ contribution, and that the OPE contribution is
small, contrarily to the bounds on $-\delta_{1/m^2}^{(A_1)}$,
$-\ell_2(1)$ and $-\delta_{1/m^2}^{(1)}$. On the other hand, the sets
of parameters (i), (iii) and (iv) are unphysical, since the lower bound
on $-\ell_1(1)$ is very large. In Table 3 we observe that the bounds on
$-\ell_2(1)$ and on $-\delta_{1/m^2}^{(1)}$ are always dominated by the
OPE contribution, except when $\rho^2$ and $\sigma^2$ approach ${3 \over 4}$ and ${15 \over 16}$, like
in the set of parameters (iii) and (iv).

\section*{Appendix C. Radiative corrections}
\hspace*{\parindent}
The radiative corrections to the relations (\ref{51e}), (\ref{52e})
have been computed by Dorsten within HQET \cite{6r}. In this approach
there are two parameters, namely the subtraction point $\mu$ and the
cut-off $\Delta$ on the sums. To avoid large logarithms, one should
take $2\Delta \cong \mu$. \par

Our relations (\ref{51e}), (\ref{52e}) are modified in the following
way (formulas (34), (35) and (18) of \cite{6r}), adopting $2\Delta =
\mu$ to simplify~:
$$\sum_{n=0}^{n(\mu /2)} \left [ \tau_{3/2}^{(2)(n)}(1) \right ]^2 = {4 \over 5} \sigma^2 (\mu ) - \rho^2 (\mu ) \left ( 1 + {32 \alpha_s \over 27 \pi}\right ) + {4 \over 5} \ {193 \alpha_s \over 675 \pi} \eqno({\rm C.1})$$

$$\sum_{n>0}^{n(\mu /2)} \left [ \xi^{(n)'}(1) \right ]^2 = {5 \over 3} \sigma^2 (\mu ) - {4 \over 3} \rho^2 (\mu ) \left ( 1 + {20 \alpha_s \over 27 \pi}\right )  - [\rho^2 (\mu)]^2 + {5 \over 3} \ {148 \alpha_s \over 675 \pi} \eqno({\rm C.2})$$

\noi Taking $\alpha_s = 0.3$ for $\mu = 2$~GeV, we obtain, keeping the algebraic factors as in (\ref{51e}), (\ref{52e})
$$\sum_{n=0}^{n(\mu /2)} \left [ \tau_{3/2}^{(2)(n)}(1) \right ]^2 = {4 \over 5} \sigma^2 (\mu ) - 1.11 \rho^2 (\mu ) + 0.02 \eqno({\rm C.3})$$

$$\sum_{n> 0}^{n(\mu /2)} \left [ \xi^{(n)'}(1) \right ]^2 = {5 \over 3} \sigma^2 (\mu ) - {4 \over 3} 1.07 \rho^2 (\mu ) - [\rho^2 (\mu)]^2 +  0.03\eqno({\rm C.4})$$

We observe that the radiative corrections do not modify in a
significant way our results, since the
corrections are small. However, we must emphasize that, using (B.4)
as a model of a relation between slope and curvature, the divergences
of the denominators of the bounds are shifted away from $\rho^2 = {3
\over 4}$ to slightly higher values.

\vskip 1 truecm 

\section*{Acknowledgement.} \hspace*{\parindent} We are indebted to the EC contract HPRN-CT-2002-00311 (EURIDICE).

\end{document}